%% file: ms.tex
\documentclass[iop,numberedappendix,appendixfloats]{emulateapj}

\usepackage{chngcntr}
\usepackage{color}

\newcommand{\J}{\textit{J}}
\newcommand{\JHKs}{\textit{JHK$_\mathrm{S}$}}
\newcommand{\Spitzer}{\textit{Spitzer}}
\newcommand{\Chandra}{\textit{Chandra}}
\newcommand{\tooc}{$\theta^1$Ori~C}

\newlength{\ECDFwidth}
\newlength{\mapheight}
\newlength{\mapwidth}
\newlength{\ninepanelwidth}
\newcommand{\ECDFcaption}{ECDFs of distances from massive stars to disk-bearing
(red line) and disk-free (blue line) YSOs. Left panel is for early O stars,
right panel is for late O stars. Dashed black line is a parabola, indicating
the predicted CDF for two-dimensional geometry.}

\newcommand{\ECDFcattwocaption}{The density of disk-bearing YSOs in this region
is too small to detect disk avoidance within the destruction radii predicted by
theory.}

\newcommand{\mapcaption}[4]{There are #1 disk-bearing YSOs (red circles), #2
disk-free YSOs (blue circles), and #3 O stars (large green circles), labeled by
spectral types from #4.}

\newcommand{\appendixcaption}{Red circles represent disk-bearing YSOs, blue
circles represent disk-free YSOs, and green circles represent O stars. The
yellow line segment corresponds with a projected linear distance of 1~pc. The
magenta lines show the \Chandra\ field of view. The cyan contours show the
8.0~\micron\ contour cutoff used to exclude heavily PAH-contaminated regions
(as discussed in \S~\ref{subsect:pahcutoffs}). Orange ellipses show the
subclusters identified by \citet{kuhnetal2014}.}

\begin{document}

\title{No evidence for protoplanetary disk destruction by OB stars in the
MYStIX sample}

\author{Alexander J.W. Richert\altaffilmark{1}, Eric D.
Feigelson\altaffilmark{1,2}, Konstantin V. Getman\altaffilmark{1} \& Michael A.
Kuhn\altaffilmark{3,4}}
\altaffiltext{1}{Dept. of Astronomy \& Astrophysics, The Pennsylvania State
University}
\altaffiltext{2}{Center for Exoplanet and Habitable Worlds}
\altaffiltext{3}{Instituto de F\'{i}sica y Astronom\'{i}a, Universidad de
Valpara\'{i}so, Gran Breta\~{n}a 1111, Playa Ancha, Valpara\'{i}so, Chile}
\altaffiltext{4}{Millennium Institute of Astrophysics}

\begin{abstract}

HST images of proplyds in the Orion Nebula, as well as submillimeter/radio
measurements, show that the dominant O7 star \tooc\ photoevaporates nearby
disks around pre-main sequence stars. Theory predicts that massive stars
photoevaporate disks within distances of order 0.1~pc. These findings suggest
that young, OB-dominated massive H~II regions are inhospitable to the survival
of protoplanetary disks, and subsequently to the formation and evolution of
planets. In the current work, we test this hypothesis using large samples of
pre-main sequence stars in 20 massive star-forming regions selected with X-ray
and infrared photometry in the MYStIX survey. Complete disk destruction would
lead to a deficit of cluster members with excess in \JHKs\ and \Spitzer/IRAC
bands in the vicinity of O stars. In four MYStIX regions containing O stars and
a sufficient surface density of disk-bearing sources to reliably test for
spatial avoidance, we find no evidence for the depletion of inner disks around
pre-main sequence stars in the vicinity of O-type stars, even very luminous
O2--O5 stars. These results suggest that massive star-forming regions are not
very hostile to the survival of protoplanetary disks and, presumably, to the
formation of planets.

\end{abstract}


\section{Introduction}
\label{sect:introduction}

Proplyds are protoplanetary disks that are heated and thereby illuminated by
ultraviolet light from nearby massive stars. The first proplyds were discovered
in the Orion Nebula's Trapezium Cluster around \tooc\ \citep[spectral type
O7;][]{simondiazetal2006} using radio observations \citep{churchwelletal1987}.
The most well-known observations of the Orion proplyds come from optical imaging
by the Hubble Space Telescope \citep{odelletal1993, odellwen1994, ballyetal1998,
smithetal2005, riccietal2008}. Many of the images show cometary tails directed
away from \tooc, which result from the combined action of far ultraviolet (FUV)
radiation, which heats and therefore expands the disk, and the stellar
wind, which creates a bow shock at the wind--disk interface
\citep{garciaarredondoetal2001}. Spectroscopic observations confirm that disks
lose mass to outflows of heated gas \citep{henneyodell1999}, which presumably
also carry along small dust particles. Indeed, sub-mm measurements of dust
emission from disks in the Trapezium Cluster reveal disk mass truncation within
tenths of a parsec of \tooc\ \citep{mannwilliams2009b, mannwilliams2010,
mannetal2014}, though in the Flame Nebula (NGC~2024), which is somewhat younger
than the Trapezium Cluster \citep{getmanetal2014}, \citet{mannetal2015} find no
evidence of such an effect around the less massive IRS~2b \citep[spectral type
O8--B2;][]{biketal2003}.

These observations have motivated models of mass loss rates for externally
photoevaporated disks as a function of incident ultraviolet flux and the
properties of the disk, which can in turn be compared with observed disk
lifetimes and initial masses \citep{johnstoneetal1998, storzerhollenbach1999,
adamsetal2004, adamsetal2006, clarke2007, fatuzzoadams2008}. Predicted mass loss
rates vary significantly among these studies; the shortest estimates of disk
lifetimes are of order $10^5$~yr \citep[e.g.,][]
{johnstoneetal1998, storzerhollenbach1999}, which agree with the
observational findings of \citet{henneyodell1999}, and are much shorter than
ordinary viscous accretion timescales of $10^6$--$10^7$~yr
\citep[][and references therein] {haischetal2001, hernandezetal2007}. By
contrast, the combined external photoevaporation and viscous accretion model of
\citet{andersonetal2013} predicts that external photoevaporation shortens disk
lifetimes by a factor of only a few, implying a minimal effect on planet
formation and evolution. Observationally constraining planet
formation and photoevaporative destruction timescales is therefore vital to
determining whether OB-dominated star-forming regions are hostile to planet
formation.

While the aforementioned theoretical works disagree somewhat as to how far
external photoevaporation by massive stars proceeds into the disk (estimates are
of order tens of AU from the disk host star), the destruction of the outer disk
may lead indirectly to the depletion of the inner disk \citep{johnstoneetal1998,
andersonetal2013}, as material accreted onto the host star is no longer
replenished by the now-missing outer disk, and accretion in the inner disk
occurs on timescales of $10^4$--$10^5$~yr \citep{hartmannetal1998}. Hence, the
total depletion of the outer disk may imply the quick disappearance of the inner
disk. By determining whether the inner regions of disks are present in the
vicinity of O stars in regions of different ages, the rate at which the
\textit{outer} regions are destroyed can be observationally constrained. This is
the goal of the present work.

A valuable new dataset that examines disk properties in a range of UV-dominated
environments is the Massive Young Star-Forming Complex Study in Infrared and
X-ray (MYStIX), a survey of 20 young, OB-dominated regions that combines X-ray
and infrared photometry to study clustered star formation, early cluster
evolution, and protoplanetary disk evolution \citep{feigelsonetal2013}. The
advantage of the MYStIX approach is that it captures both X-ray selected
disk-free and infrared-selected disk-bearing stars. The catalog of MYStIX
Probable Complex Members (MPCMs) consists of 31,784 young stellar objects
\citep[YSOs;][]{broosetal2013}. Disk-bearing YSOs are identified by infrared
excess in the 1--8\micron\ spectral energy distribution (SED), while disk-free
YSOs are identified through X-ray emission and an SED in the 1--8\micron\ range
that is consistent with a bare stellar photosphere. Membership criteria and YSO
classification are discussed in Sections \ref{subsect:membership} and
\ref{subsect:classification}, respectively. The catalog also includes OB stars
already identified in the literature (discussed further in
\S~\ref{subsect:obselection}).

The MPCM catalog, with thousands of lower-mass YSOs and dozens of OB stars,
therefore in principle contains the information needed to determine whether
external photoevaporation by massive stars leads to the complete depletion of
the inner disk. Celestial coordinates from the MPCM catalog, along with known
region distances (ranging 0.4--3.6~kpc), can be used to calculate projected
distances between disk-bearing YSOs and O stars in each MYStIX region. With
this information, we are able to test for any tendency of disk-bearing YSOs to
spatially avoid O stars, which would imply complete photoevaporative
destruction of the disks. It should be emphasized that a failure to find
evidence that O stars are spatially avoided by disk-bearing YSOs would not
imply that disks are not externally photoevaporated; rather, it would mean that
the process of external photoevaporation is not as rapid or complete as
predicted by some theory.

After examining all 20 MYStIX regions, we find no evidence of disk
destruction\footnote{For the remainder of the paper, unless otherwise
specified, the term ``disk destruction" will refer to the \textit{complete}
depletion of the disk due to external photoevaporation, possibly aided by
viscous accretion.}, with the results for two regions (Carina Nebula and M~17)
being inconclusive due to observational difficulties. It seems that the
ultraviolet radiation fields found in OB-dominated regions do not present such
a hostile environment for protoplanetary disks as suggested in some previous
works. In Section~\ref{sect:methods}, we discuss membership selection, YSO
classification, and the exclusion of high-nebulosity regions in each MYStIX
field. In Section~\ref{sect:results}, we present the results of our tests for
spatial avoidance of O stars by disk-bearing YSOs, and compare these results
with several previous works for two regions. Interpretation relating to disk
destruction astrophysics and implications for planet formation are discussed in
Section~\ref{sect:conclusions}.

\section{Methods}
\label{sect:methods}

\subsection{SFR membership}
\label{subsect:membership}

The MYStIX catalog is derived by cross-matching photometric point sources from
X-ray, near-infrared, and mid-infrared bands. X-ray archival data are taken
from the ACIS instrument on the \textit{Chandra X-ray Observatory}; near-IR
(\JHKs) data for each region come mostly from either 2MASS or United Kingdom
Infrared Telescope (UKIRT) archival data; and mid-infrared photometry is taken
from the \textit{Spitzer Space Telescope}'s IRAC instrument (3.6~\micron,
4.5~\micron, 5.8~\micron, and 8.0~\micron). Further details for these data and
their analysis are given by \citet{feigelsonetal2013} and the associated MYStIX
observational papers \citep{kuhnetal2013a, kuhnetal2013b, nayloretal2013,
povichetal2013, broosetal2013, townsleyetal2014}.

\citet{broosetal2013} assign a probability of region membership to each X-ray
source using a naive Bayes classifier, wherein each attribute is assumed to
contribute independently to the probability of membership. Attributes used for
this classification are \J-band magnitude; X-ray median energy, variability, and
local source density; 4.5~\micron\ magnitude; infrared excess (with respect to
bare stellar photosphere); and mid-IR local source density. Each X-ray source is
subsequently classified as a foreground star, an MPCM, a background (Galactic)
star, or a galaxy/active galactic nucleus, using training sets constructed
according to the methods presented by \citet{getmanetal2011} and
\cite{broosetal2013}.

Infrared and X-ray sensitivities vary between MYStIX regions due to different
telescope exposures, region distances, and levels of absorption. X-ray
sensitivity also varies \textit{within} each MYStIX field. It is therefore
difficult to reliably compare the surface densities of YSOs of each class among
different MYStIX regions. However, this does not present a significant problem
for studying YSO surface densities in the vicinity of O stars across the
relatively small distances (tenths of a parsec) involved in disk destruction.

In each region, completeness with \J-band magnitude varies, and certainly many
low-mass members are missing from the MYStIX sample. This limits the scope of
our conclusions, as disks around lower-mass stars are predicted to experience
higher mass-loss rates from external photoevaporation due to lower escape speed
for outflows of heated gas \citep[see, e.g.,][]{andersonetal2013}. A failure to
identify disk destruction for the stars in the MYStIX sample therefore does not
preclude the possibility that the disks around stars with masses below our
completeness limits are being completely destroyed.

\subsection{YSO classification}
\label{subsect:classification}

\citet{povichetal2013} classify MPCMs by YSO class using both color-color cuts
(to exclude contaminants and bad photometry from diffuse nebular emission) and
fitting of model spectral energy distributions (SEDs) based on near- and
mid-infrared photometry for each source. The stellar atmosphere models of
\citet{castellikurucz2004} and the YSO SED models of \citet{robitailleetal2006}
are used. Observed near-/mid-IR SEDs most consistent with bare stellar
photospheres are classified as disk-free (Class III YSO), while those consistent
with the model SEDs of Class I or II YSOs are classified as disk-bearing.
Disk-free YSOs are additionally required to be detected as X-ray sources in
order to avoid contamination by the often very populous field star population.

For 18 of 20 MYStIX regions, in order for a given source to be classified, it
must be detected at 3.6~\micron\ and 4.5~\micron, in addition to at least two
of the remaining five bands: \JHKs, 5.8~\micron, and 8.0~\micron. The Carina
Nebula was analyzed in a separate work \citep{povichetal2011}, where photometry
for \textit{any} four of the seven infrared bands were required. For the Orion
Nebula, we use the catalog of \citet{megeathetal2012}, who classified YSOs
based on mid-infrared colors and spatial distribution, and, as in the case of
Carina, require that disk-free YSOs be detected in 4 of 7 infrared bands.

Although the MYStIX sample includes both disk-bearing and disk-free stars, it
should be noted that the science goals of this paper can in principle be met
using only data for disk-bearing YSOs (along with the O stars). We test for disk
destruction by looking for any apparent dropoff in the surface density of disks
found at small distances from O stars. We do not calculate disk fraction as a
function of distance from O stars, because the estimation of ratios of
Poisson-distributed variables can be very uncertain \citep{brownetal2001,
parketal2006}. We also include a parallel analysis of disk-free YSOs as
a test of reliability for any apparent avoidance effect for the disks. The disk
destruction interpretation of an apparent tendency for disk-bearing YSOs to
avoid O stars would be undermined by any tendency for disk-free YSOs to do the
same, which would imply observational rather than astrophysical origins for the
avoidance effect.

\subsection{Selection of OB stars}
\label{subsect:obselection}

The MYStIX catalog incorporates published OB stars based on optical
spectroscopy. Spectral types are taken from \citet{skiff2009} and SIMBAD, and
range from B3 to O2. Positions from the literature are matched to \JHKs\
photometry as described by \citet{broosetal2013}.

\subsection{Excluding PAH-contaminated regions}
\label{subsect:pahcutoffs}

Many MYStIX regions contain extensive nebulosity where ultraviolet radiation
and winds from OB stars heat dust on the periphery of the surrounding molecular
cloud, leading to thermal emission from polycyclic aromatic hydrocarbons (PAHs)
at mid-infrared wavelengths (particularly in the IRAC 5.8~\micron\ and
8.0~\micron\ bands). In principle, mid-IR source contamination at these
wavelengths due to bright PAH emission could increase the apparent disk
fraction in a given area on the sky, since disk-free sources will be more
difficult to detect in the IR bands. A disk-free source in a PAH-dense region
is also more likely to be misclassified as disk-bearing than a disk-bearing one
is to be misclassified as disk-free, which again will lead to an overestimate
of disk fraction. Depending on the position of such a region with respect to O
stars, a disk destruction effect could potentially be falsely created or
falsely obscured.

\begin{figure}
\centering
\includegraphics[width=0.475\textwidth]{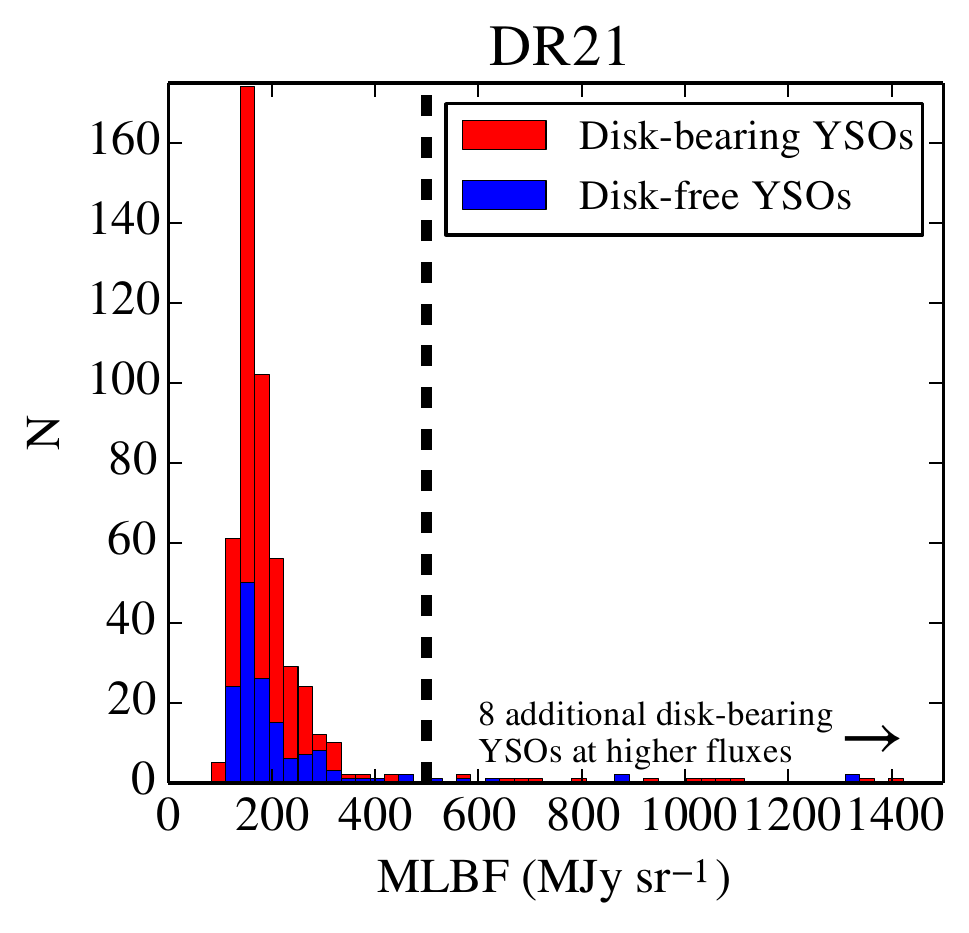}
\caption{Histogram of median local 8.0~\micron\ background flux (MLBF) around
disk-bearing and disk-free YSOs in DR~21. Vertical line shows the chosen MLBF
level above which stars are excluded from our analysis.}
\label{fig:dr21_fluxhist}
\end{figure}

To mitigate the problem of PAH contamination, we exclude the most strongly
contaminated areas of each of the 20 MYStIX regions based on 8.0~\micron\
\Spitzer/IRAC images, which also excludes those regions where infrared images
are strongly contaminated by the PSFs of very bright stars, such as $\eta$
Carinae (Carina Nebula). In each region, we calculate the median 8.0~\micron\
flux around each YSO within a $\sim$7"$\times$7" square. We construct
histograms of these median local background fluxes (MLBFs) for disk-bearing and
disk-free YSOs in each MYStIX region. Figure~\ref{fig:dr21_fluxhist} shows this
analysis for a typical case, DR~21, where the disk-bearing and disk-free
distribution follow each other up to an MLBF of $\sim500$~MJy~sr$^{-1}$. At
higher MLBFs, more disks are found due to their higher mid-infrared fluxes, as
well as possible misclassifications due to unreliable photometry in PAH-dense
regions. An MLBF of 500~MJy~sr$^{-1}$ threshold for the DR~21 field is also
validated by visual examination of the 5.6 and 8.0~\micron\ images; fainter
point sources are more difficult to identify at this and brighter nebulosity
levels. In each region, at the MLBF where the two distributions seem to diverge
in this way, we select a cutoff value. We then calculate contours based on this
cutoff value in the 8.0~\micron\ \Spitzer/IRAC image for each region, and
exclude all sources (including O stars) from our analysis. Possible
complications relating to PAH regions and their exclusion are discussed
throughout Sections \ref{sect:results} and \ref{sect:conclusions}.

\subsection{Summary of membership}
\label{subsect:regionsummary}

For each MYStIX region, Table~\ref{table:regiondata} shows distances, selected
MLBF cutoff levels, and the numbers of O stars ($N_\mathrm{O}$),
disk-bearing YSOs ($N_\mathrm{D}$), and disk-free YSOs
($N_\mathrm{ND}$) in the MPCM sample after MLBF cutoffs are applied.
For disk-bearing YSOs in each region, we also compute the approximate 50\%
mass completeness limits, $\mathcal{M}_{50}$, in other words, 50\% of stars of
mass $\mathcal{M}_{50}$ are detected in our sample, based on a fit of the
stellar initial mass function of the form given by \citet{maschberger2013}.
Regions are divided into three categories, described in
Section~\ref{sect:results}. Information on subcluster ages and structures can be
found in \citet{getmanetal2014} and \citet{kuhnetal2014}, respectively.

\input{t1.tex} %\label{table:regiondata}

\section{Results}
\label{sect:results}

\begin{figure*}
\centering
\includegraphics[width=0.9\textwidth]{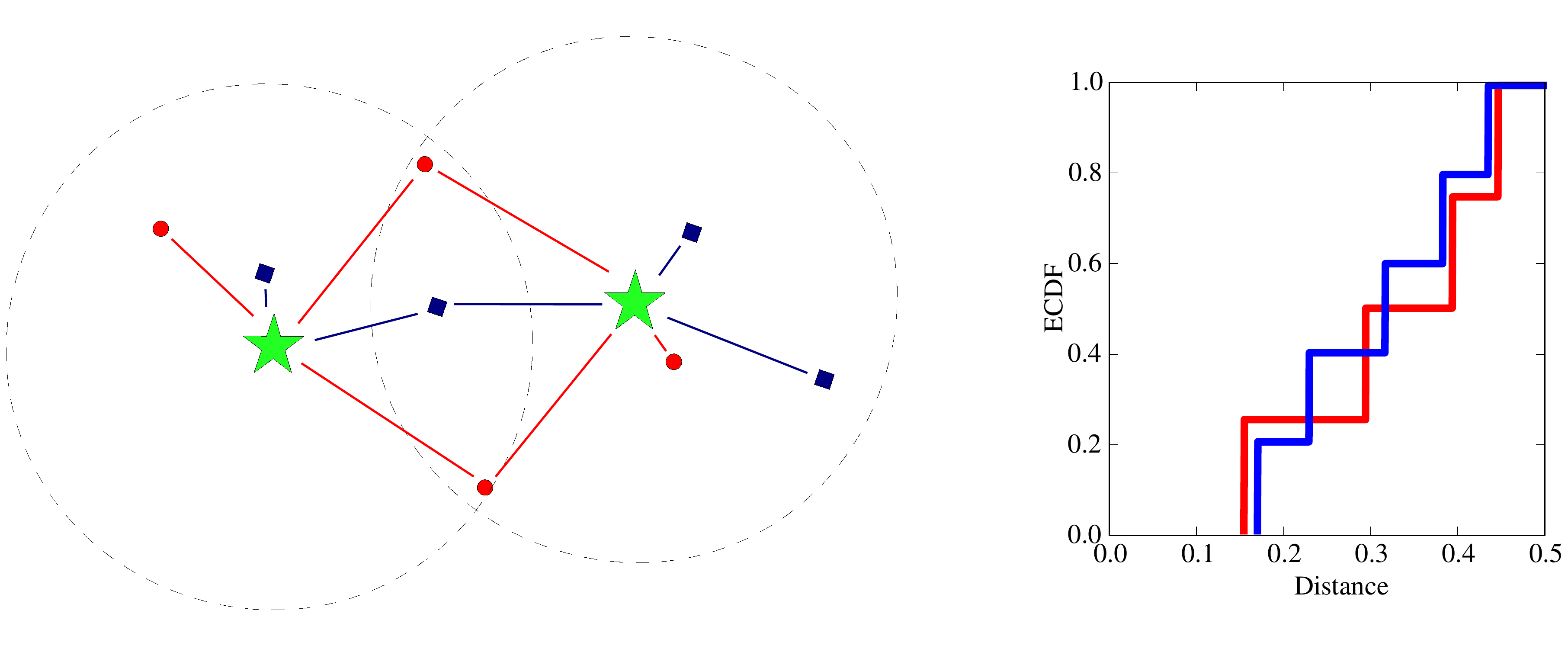}
\caption{Illustration of distance ECDF scheme used in the current work using
mock data. Left: Green stars represent O stars, red circles are disk-bearing
YSOs, and blue squares are disk-free YSOs. Dashed circles represent 0.5~pc
radius from each O star. Right: ECDF of O star--YSO distances based on left
panel, only including distances under 0.5~pc.}
\label{fig:cartoonECDF}
\end{figure*}

In order to test for disk destruction in a given region, we calculate projected
distances between O stars and their disk-bearing and disk-free neighbors. For
each region under study, we combine these data separately for early- and
late-type O stars and construct empirical cumulative distribution functions
(ECDFs) for distances within 0.5~pc, a limiting distance which is well beyond
the radii of disk destruction predicted by theory\footnote{Extending the ECDF
analysis to greater distances (say, 1~pc rather than 0.5~pc) does not affect
our results. The use of 0.5~pc as the limiting distance does, however, make the
visual comparison of distances to disk-bearing and disk-free YSOs easier. At
these short distances the distributions are nearly parabolic, while at larger
distances this parabolic shape breaks down due to the non-uniform structure of
the regions at these scales, as well as gradients in both disk fraction (e.g.,
due to the presence of nearby protostar-dominated embedded clusters) and
instrumental sensitivities.}. Figure~\ref{fig:cartoonECDF} illustrates the
construction of these ECDFs using mock data. Disk destruction would be revealed
by a sudden dearth of disk-bearing YSOs at short distances from O stars. The
use of ECDFs avoids arbitrary radial binning, and the separate analysis of
disk-bearing and disk-free YSOs avoids uncertain small-sample disk fractions.

The observed geometries of the regions under study are complex due to limited
X-ray fields of view, PAH-contaminated regions and the exclusion thereof, and
the intrinsic concentration of YSOs into distinct subclusters with different
ages \citep{kuhnetal2014, getmanetal2014}. Therefore, it cannot be assumed
\textit{a priori} that the number of YSOs around a given O star will increase
with distance in a predictable manner. In practice however, this distribution is
close to parabolic within distances of $\sim$0.5~pc, as seen in the figures in
Sections \ref{subsect:cat2} and \ref{subsect:cat3}. As such, for each ECDF
result, we also plot a uniform distribution based on two-dimensional geometry,
i.e., a parabola, while also taking into account the effect of PAH exclusion
regions and the edge of the X-ray field of view. These parabolas provide a test
of statistical significance for any apparent avoidance effect. The presence of
fewer disk-bearing YSOs than disk-free within some distance of O stars is not
regarded as evidence for disk destruction so long as the ECDF for the
disk-bearing YSOs follows its predicted parabolic increase. This is because the
distance between an O star and the nearest disk-bearing YSO is expected to be
large in a region with an intrinsically sparse population of disks.

We organize the results for each of the 20 MYStIX regions into one of three
categories: those without O stars (\S~\ref{subsect:cat1}), those with O stars
but with apparent surface densities of disk-bearing YSOs that are too low to
test for avoidance of O stars (\S~\ref{subsect:cat2}), and those with O stars
\textit{and} sufficiently dense disk-bearing YSOs to test for disk destruction
(\S~\ref{subsect:cat3}). The term ``sufficiently dense" will be explained in
Section~\ref{subsect:cat2}. Appendix~\ref{sect:app1} contains figures showing,
for each MYStIX region containing O stars, the spatial distribution of O stars,
disk-bearing and disk-free pre-main sequence stars, and excluded areas with
nebular contamination above the MLBF cutoff superposed on the \Spitzer/IRAC
8.0~\micron\ image.

\subsection{Regions without O stars}
\label{subsect:cat1}

Seven of 20 MYStIX regions lack O stars in the sample used in this work. DR~21
has no O stars to begin with, with the only OB-type star in the MYStIX catalog
being a B0.5V star. The other six regions each have one or two O stars, but all
were excluded from our analysis due to the PAH exclusion scheme described in
Section~\ref{subsect:pahcutoffs}. These are the Flame Nebula, NGC~2362, RCW~36,
RCW~38, the Trifid Nebula, and W~40. Except for RCW~38, these regions contain
only one or two late O stars in the non-PAH excluded sample, and are therefore
unlikely to produce a disk destruction effect. In some cases, the surrounding
density of disk-bearing YSOs (to the extent that they can be observed on
account of PAH contamination) are so low that disk destruction would not occur
anyway, as in the regions described below in Section~\ref{subsect:cat2}. RCW~38
contains an O5.5V star, however the apparent surface density of YSOs around it
is far too low for significant disk destruction to be observed.

\subsection{Regions with low apparent densities of disk-bearing YSOs}
\label{subsect:cat2}

In order to determine whether disks spatially avoid O stars within some radius
due to complete photoevaporative destruction, the surface density of
disk-bearing YSOs must be great enough to detect a deficit of them at short
distances to O stars. Astrophysically, disk destruction cannot occur if there
are no disks present in the vicinity of a given O star, even if the UV radiation
field within some radius of the star is in principle high enough to
photoevaporate a disk. Observationally, disk destruction cannot be detected if
too few disk-bearing YSOs are present to reliably measure the distribution of
their distances from O stars and detect a dropoff at short distances.

In 7 out of the 13 MYStIX regions containing O stars, the observed surface
densities of disk-bearing YSOs around O stars are too small to detect
destruction of nearby ($d\lesssim0.2$~pc) disks: the Lagoon Nebula, NGC~1893,
NGC~2264, NGC~3576, NGC~6334, W~3, and W~4. These regions therefore cannot be
used to constrain the physics of disk destruction. ECDF results for these seven
regions are shown in Appendix~\ref{sect:app2}.

\citet{kuhnetal2015} estimate intrinsic YSO surface densities for 17 of
20 MYStIX regions, including four of the seven regions discussed in this
section. We determine the local intrinsic surface densities (LISDs) for the
eight O stars in those four regions, and compare them with the LISDs of the O
stars in regions with higher apparent surface densities
(\S~\ref{subsect:cat3}). We confirm that LISDs around O stars in the four
regions discussed in this section are systematically lower, suggesting that
intrinsically low surface densities, convolved with relatively high stellar
mass sensitivity limits of $\ge0.6$~$M_\odot$, are responsible for the sparsely
populated ECDFs seen in Figures~\ref{fig:app2A}--\ref{fig:app2B}.

It is interesting to note that in the case of NGC~2264, there is a relatively
large number of disk-bearing YSOs located within 0.5~pc of the late O stars,
however a strong disk fraction gradient in the region leads to significantly
non-parabolic distributions for both the disk-bearing and disk-free YSOs due to
the global age gradient in NGC~2264 identified by \citep{getmanetal2014}. This
prevents the identification of disk destruction at distances less than 0.2~pc.

\subsection{Regions tested for disk destruction}
\label{subsect:cat3}

Six of 20 MYStIX regions have O stars and a sufficient density of disk-bearing
YSOs to test for disk destruction: M~17, NGC~6357, and the Carina, Eagle, Orion,
and Rosette Nebulae. In all six cases, our ECDF analysis reveals no evidence
that disk-bearing YSOs spatially avoid O stars in a manner consistent with
photoevaporative dissipation. In the following, we discuss each of these
regions. For the Eagle and Rosette Nebulae, we also compare our results with
previous similar works examining disk fraction around massive stars based on
infrared excesses \citep{guarcelloetal2007, guarcelloetal2009, balogetal2007}.

\subsubsection{Carina Nebula}
\label{subsubsect:carina}

\begin{figure*}
\centering
\includegraphics[width=\ECDFwidth]{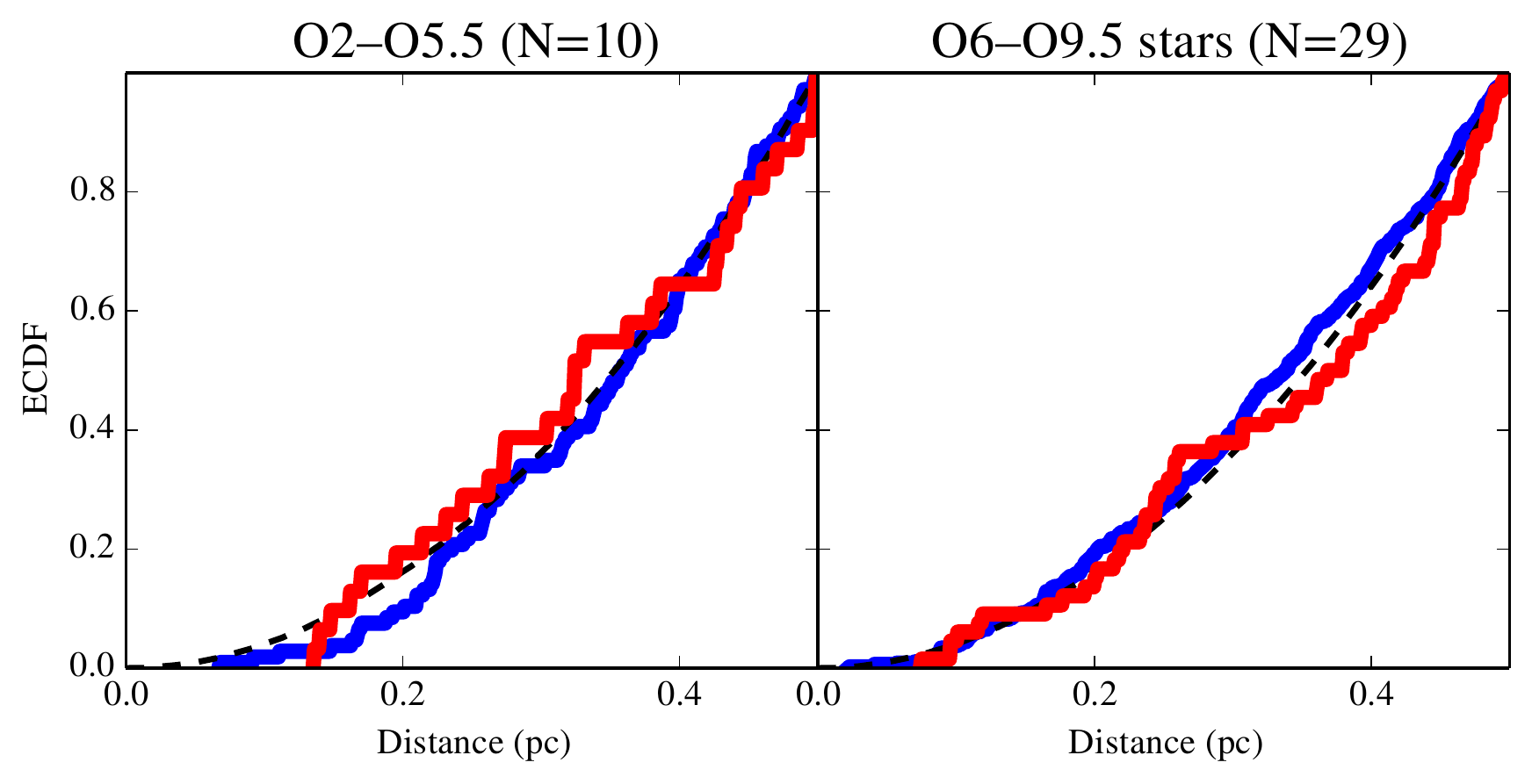}
\caption{Carina Nebula. \ECDFcaption}
\label{fig:carina_ECDF}
\end{figure*}

ECDF results for the Carina Nebula are shown in Figure~\ref{fig:carina_ECDF},
with the caveat that we have excluded O stars in Trumpler~14, a subregion of
Carina which suffers from significant crowding, resulting in unreliable infrared
photometry. Figure~\ref{fig:carina_map} shows Trumpler~14 to the northwest next
to an illuminated molecular cloud above the MLBF cutoff.

Though the disk-bearing YSOs of Carina appear to avoid early O stars at
distances of $\sim$0.1~pc, the disk-free YSOs also appear underabundant at
distances up to $\sim$0.15~pc. Performing this analysis without the requirement
of detection in four infrared bands for the disk-free objects (i.e.,
designating all non-infrared excess sources as disk-free), we find that this
apparent avoidance effect for the disk-free objects disappears. This strongly
suggests that the origins of this effect are observational rather than
astrophysical.

The Carina Nebula is 2.3~kpc away; at this distance, a linear separation of
0.1~pc between two objects corresponds with an angular separation of just under
10 arcsec, which is comparable to the observed point spread functions (PSFs) of
early O stars in several infrared bands. This problem of crowding may be further
exacerbated by local illumination of PAHs below the MLBF cutoff, extending the
effective PSF of each O star in the mid-infrared.

Our results for the Carina Nebula are therefore inconclusive, and would best be
studied through multiwavelength observations at higher resolutions to overcome
the problem of crowding. For now, the hypothesis of disk destruction within a
few tenths of a parsec of early O stars in the Carina Nebula cannot be
conclusively supported or refuted.

\subsubsection{M~17}
\label{subsubsect:m17}

\begin{figure*}
\centering
\includegraphics[width=\ECDFwidth]{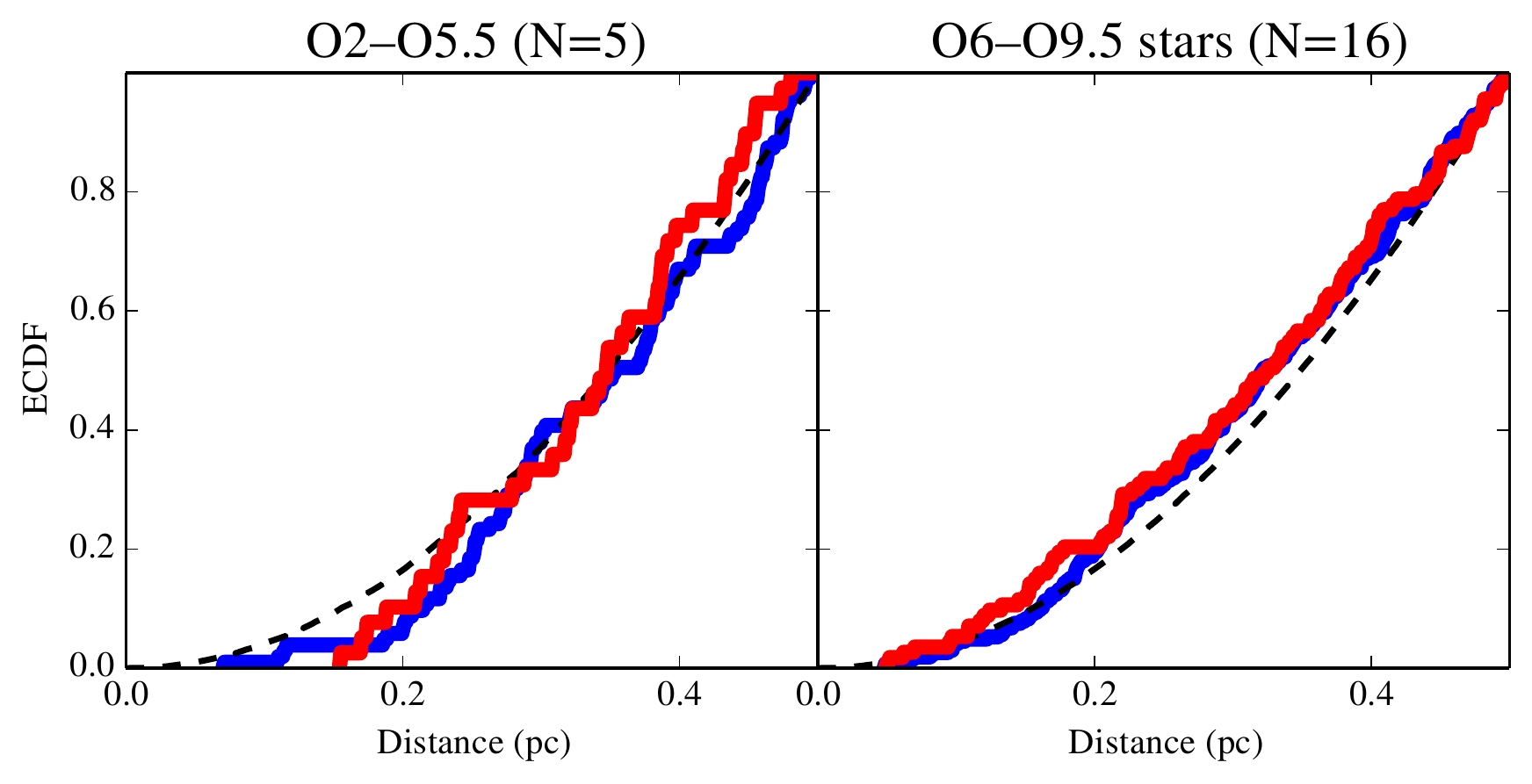}
\caption{M~17. \ECDFcaption}
\label{fig:m17_ECDF}
\end{figure*}

Figure~\ref{fig:m17_ECDF} shows results for M~17. As in the Carina Nebula,
disk-bearing YSOs seem to spatially avoid early O stars within $\sim$0.15~pc,
however the disk-free YSOs appear to be underabundant at these distances as
well. As with the aforementioned case of Carina, an astrophysical explanation
is unlikely. Instead, a combination of crowding, large PSFs of early O stars,
and PAH contamination result in a decreased detection of all infrared stars in
the \Spitzer\ images in the dense core of M~17. Less than 25\% of MYStIX X-ray
sources in M~17 have reliable infrared photometry in enough bands for SED
classification, suggesting that crowding plays an important role. As with
Carina, disk destruction is neither supported nor refuted by the available
data.

\subsubsection{Eagle Nebula}
\label{subsubsect:eagle}

\begin{figure*}
\centering
\includegraphics[width=\ECDFwidth]{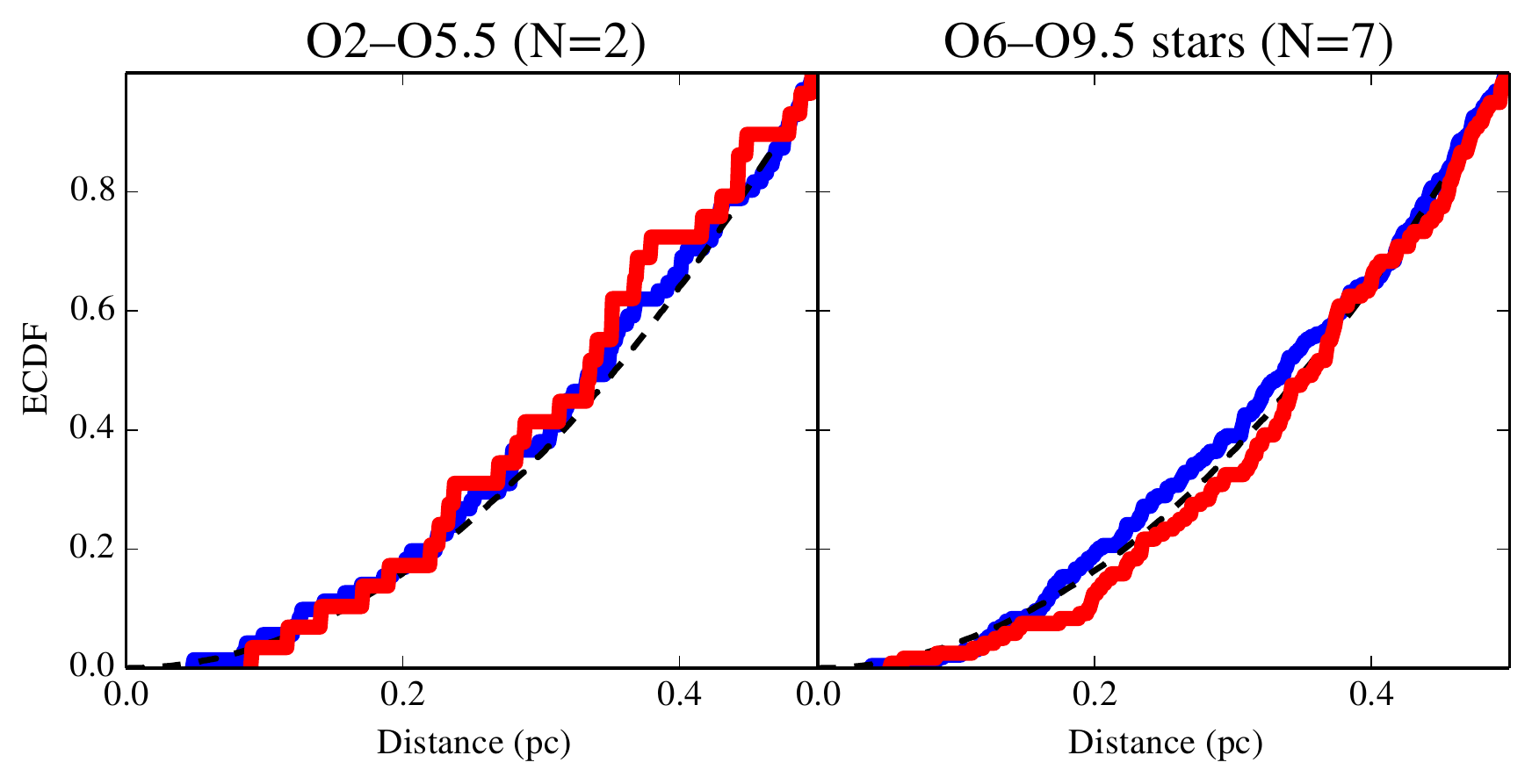}
\caption{Eagle Nebula. \ECDFcaption}
\label{fig:eagle_ECDF}
\end{figure*}

Figure~\ref{fig:eagle_ECDF} shows results for the Eagle Nebula. For both early
and late O stars, disk-bearing and disk-free YSOs follow the parabolic
distributions expected for two-dimensional geometry, and the disk-bearing YSOs
show no sign of O star avoidance. This result conflicts with the findings of
\citet{guarcelloetal2007} and \citet{guarcelloetal2009}, who report a decrease
in disk fraction for the YSOs most strongly illuminated by O and early B stars
in NGC~6611, the central dense region of the Eagle Nebula. In the following
paragraphs, we identify the sources of this discrepancy.

\citet{guarcelloetal2007} calculate total incident bolometric flux from massive
stars (spectral type earlier than B5) on each disk-bearing and disk-free YSO in
NGC~6611, based on the OB spectral types of
\citet{hillenbrandetal1993}\footnote{Calculating incident ultraviolet fluxes
rather than incident bolometric fluxes would arguably be more physically
justified for studying external photoevaporation of disks by massive stars, but
this approach may still in principle be capable of revealing a destruction
effect.}. The YSO membership catalog was based initially on \textit{BVI}, 2MASS,
and \Chandra/ACIS data, with disks identified from color-color diagrams (as
opposed to the model SED fitting used in the current work). The catalog and YSO
classifications were updated by \citet[][hereafter G09]{guarcelloetal2009} using
\Spitzer/IRAC data. In both papers, the authors find a significant decrease in
disk fraction of the most illuminated YSOs, suggesting that some disks have been
photoevaporated by massive stars.

In order to investigate the apparent disagreement between the results of G09
and the current work, we apply our ECDF-based analysis to the data of G09, and
apply the bolometric flux-based analysis of G09 to MYStIX data within NGC~6611.
Incident bolometric fluxes from massive stars in the same range of spectral
type (O--B4.5) were calculated using two-dimensional projected distances.
Bolometric luminosities were determined from spectral type; for O stars, we use
the synthetic photometry of \citet{martinsplez2006}, and for B stars, we use
the values given by \citet[][Appendix G]{carrollostlie2006}.

\begin{figure*}
\centering
\includegraphics[angle=90,width=\ninepanelwidth]{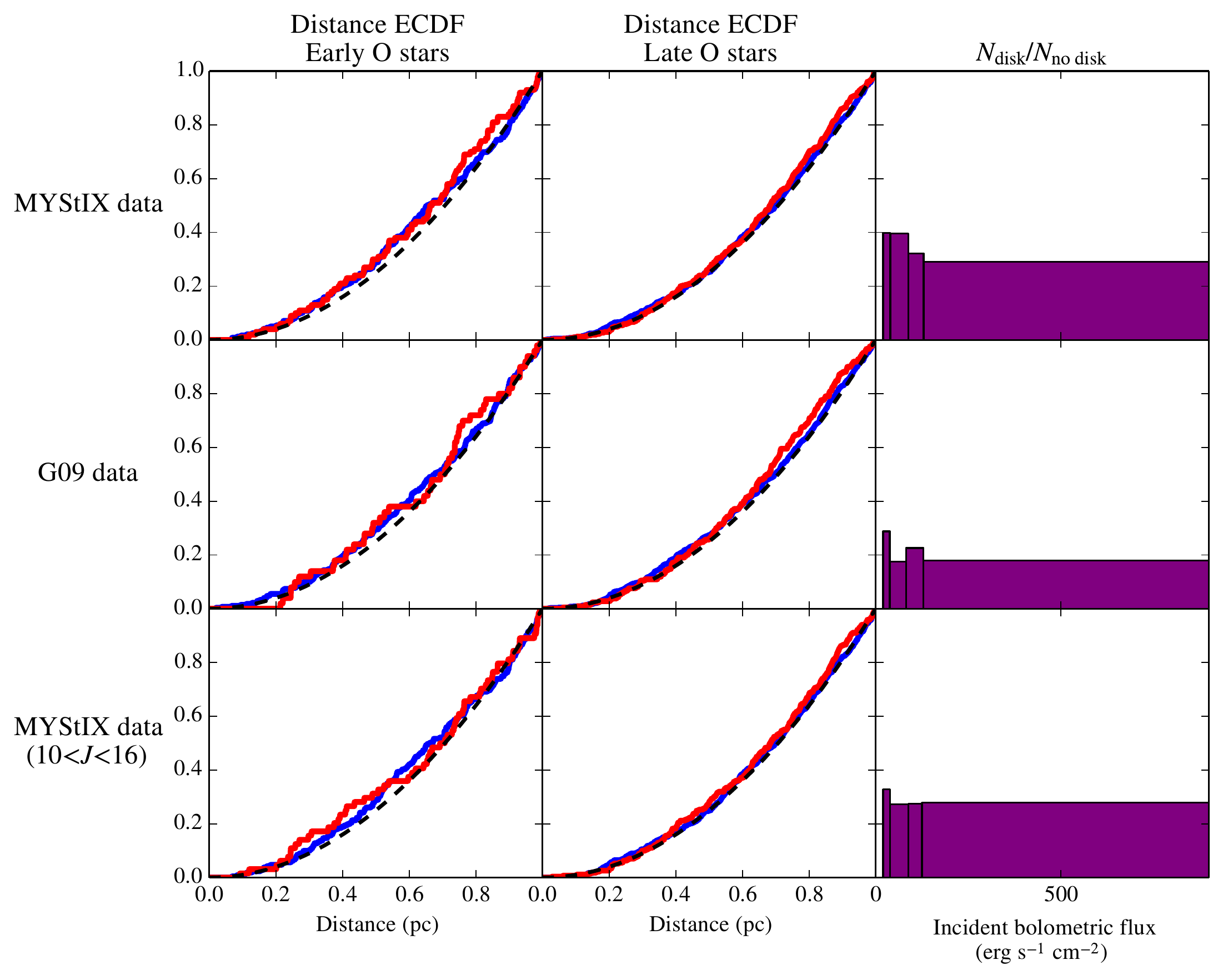}
\caption{Comparison of the data and methods of \citet{guarcelloetal2009} and
the current work. Differences in samples of disk-bearing YSOs appear to explain
the apparent disk destruction effect observed by G09. Results for disk fraction
as a function of incident bolometric flux (right three panels) are strongly
sensitive to binning scheme used, particularly the minimum incident flux value
considered (in this case, 20~MJy~sr$^{-1}$, as in G09).}
\label{fig:ninepanels}
\end{figure*}

Figure~\ref{fig:ninepanels} shows the results of these analyses. The upper three
panels show the analysis of MYStIX data using distance ECDFs for early/late O
stars and calculating disk fraction for bins of incident bolometric flux as in
G09. The middle row shows the same set of methods applied to the data of G09. We
also consider the possibility of differing mass completeness limits between
disk-bearing and disk-free YSOs within the MYStIX and G09 samples. We find that
the \J-band luminosity functions of the disk-bearing and disk-free members are
similar for the G09 sample, whereas they differ significantly in the MYStIX
sample. We therefore explore the effect of truncating the MYStIX sample by
\J-band magnitude, imposing a cut of 10$<$\J$<$16. Results are shown in the
lower three panels.

The left three panels of Figure~\ref{fig:ninepanels} show that the difference
between our results and those of G09 is primarily due to membership. The MYStIX
sample within NGC~6611 contains many more disk-bearing members than the G09
sample (715 in MYStIX sample versus 264 in G09 sample). This is likely due to
the fact that for near-IR photometry, MYStIX uses the UKIDSS survey, while G09
uses 2MASS, whose limiting \JHKs\ magnitudes are not as deep.

\begin{figure*}
\centering
\includegraphics[height=\mapheight]{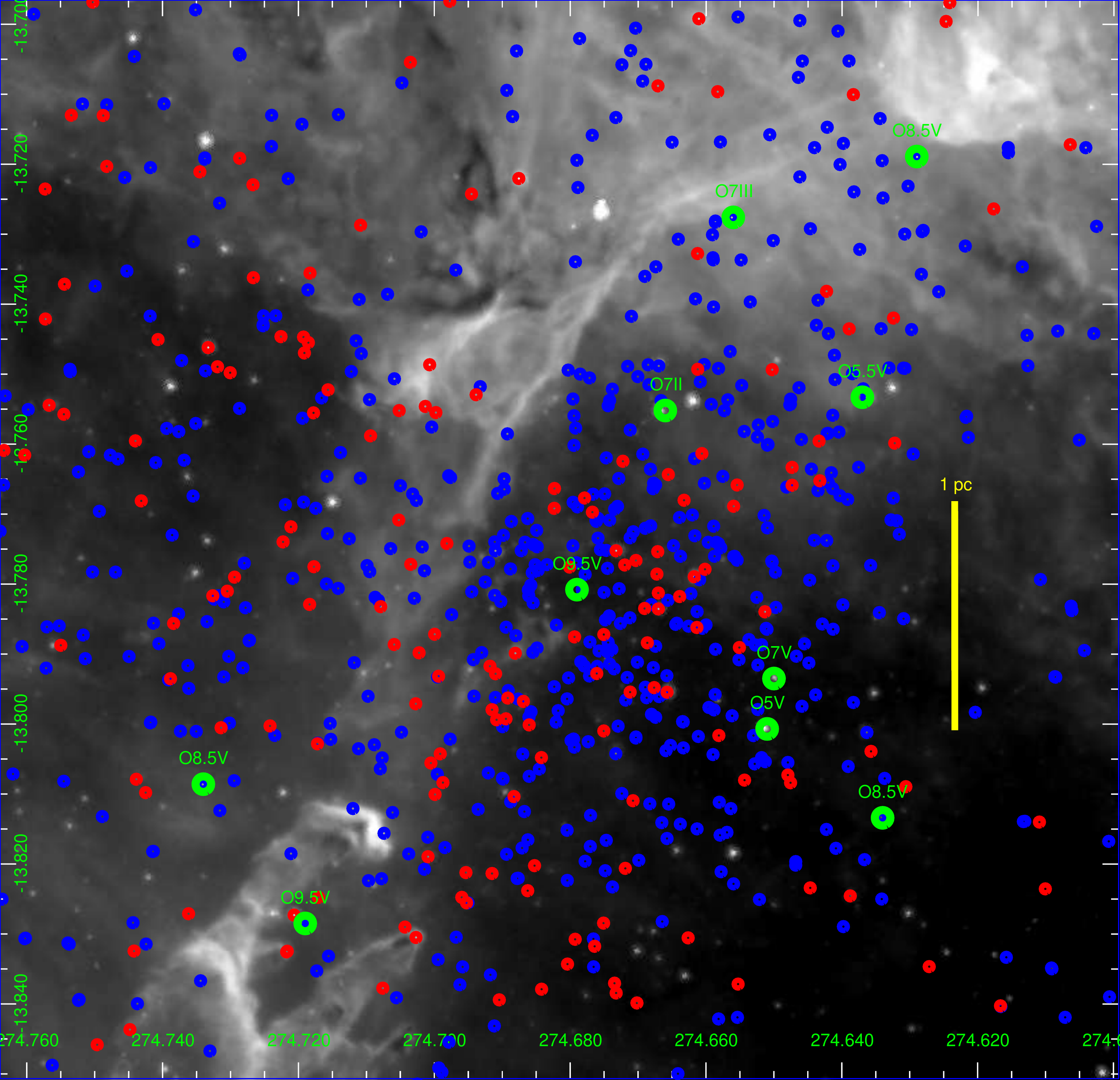}
\caption{RA--Dec map of NGC~6611 (Eagle Nebula), \textbf{using data from
\citet{guarcelloetal2009}} and 8.0~$\micron$ \Spitzer/IRAC image.
\mapcaption{264}{759}{10}{\citet{hillenbrandetal1993}} The two earliest O stars
(O5V and O5.5V) lie toward the edge of the cluster where the disk fraction is
lower, suggesting that the destruction effect reported by G09 is instead the
result of disk fraction gradients in the cluster, as well as incompleteness of
the sample of disk-bearing YSOs.}
\label{fig:guarcello_map}
\end{figure*}

Of the six panels of Figure~\ref{fig:ninepanels} showing ECDF-based analysis,
only one shows an apparent destruction effect, which is for the two early O
stars using the G09 YSO sample. Figure~\ref{fig:guarcello_map} is an RA--Dec map
of NGC~6611, showing the O stars along with disk-bearing and disk-free YSOs of
the G09 sample. Both of the early O stars, which dominate the high energy flux
in the region, are located toward the edge of the cluster, where especially for
the region to the northwest (upper right), the disk fraction appears
significantly lower even in areas without O stars. In the case of the O5V star
toward the southwest (lower right), MYStIX identifies at least half a dozen
disk-bearing YSOs not seen in the G09 sample, which eliminates the avoidance
effect.

The right three panels of Figure~\ref{fig:ninepanels} show disk fraction as a
binned function of incident bolometric flux from OB stars (hereafter simply
``flux"). We have used the same range in flux used by G09
(20--900~erg~s$^{-1}$~cm$^{-2}$). Interestingly, only the MYStIX data show a
monotonic decrease in disk fraction with flux similar to that seen in the
Figure~13 of G09. The difference between our flux analysis of the G09 data
(right column, middle row) and Figure~13 of G09 likely stems from the use of
different bolometric luminosities for OB stars. G09 does not discuss the
calculation of these quantities. Nonetheless, the flux values fall within a
similar range, and in the following paragraph we discuss the sensitivity of the
binned flux distributions to arbitrary choices made in the G09 analysis, which
may further explain this discrepancy.

In order to assess the reliability of the binned flux-based analysis of G09, we
test a range of values for the lower and upper bin edges (in G09 and
Figure~\ref{fig:ninepanels}, these are 20 and 900~erg~s$^{-1}$~cm$^{-2}$,
respectively). While the shape of the distribution is not sensitive to the
choice of upper limit, we find that it is strongly sensitive to the choice of
lower limit. Depending on the value chosen, an apparent destruction effect can
be artificially produced or made to disappear completely with the same data. If
a lowered disk fraction in the highest flux bin is to be interpreted as the
result of disk destruction, then the overall shape of the distribution should
not depend on the arbitrary choice of lower flux cutoff.

In our reanalysis of the G09 data set using the G09 flux histograms, if no
lower flux cutoff is imposed, then a steep dropoff in disk fraction with
incident flux is observed, as in the original G09 result. For the MYStIX data
set, Figure~\ref{fig:ninepanels} shows a steep drop-off in the incident flux
distribution, however this apparent destruction effect disappears completely if
a lower flux cutoff of 50~erg~s$^{-1}$~cm$^{-2}$ is applied. Of the G09 point
sources with incident fluxes less than 50~erg~s$^{-1}$~cm$^{-2}$, about a third
of them are located at projected distances of more than a parsec from the
nearest B star, and are typically even farther from the nearest O star. It would
therefore seem that high disk fractions on the periphery of the G09 field of
view, which are apparent in Figure~\ref{fig:guarcello_map}, dominate the shape of
the distribution even in the high incident flux bin. The disk fraction in these
outlying regions is not relevant to disk destruction, and the size of the field
of view under study, which is arbitrary in an astrophysical sense, should not
dictate the outcome of a test for disk destruction among YSOs located near
massive stars.

We conclude that the apparent disk destruction effect reported by G09 is not
astrophysical, and instead results from sample incompleteness and the
choice of binning scheme used in the G09 analysis, which is strongly sensitive
to disk fractions at very large distances from massive stars. It remains
unclear whether the analysis of incident bolometric fluxes by G09 is in
principle capable of detecting disk destruction. We believe that the
non-detection of disk destruction in NGC~6611 in the current work is more
reliable, as the MYStIX sample of disk-bearing YSOs is significantly more
complete (including in the vicinity of the earliest O star in the region), and
our ECDF-based analysis avoids arbitrary binning schemes and is less sensitive
to the surface densities of YSOs at large distances.

\subsubsection{Rosette Nebula}
\label{subsubsect:rosette}

\begin{figure*}
\centering
\includegraphics[width=\ECDFwidth]{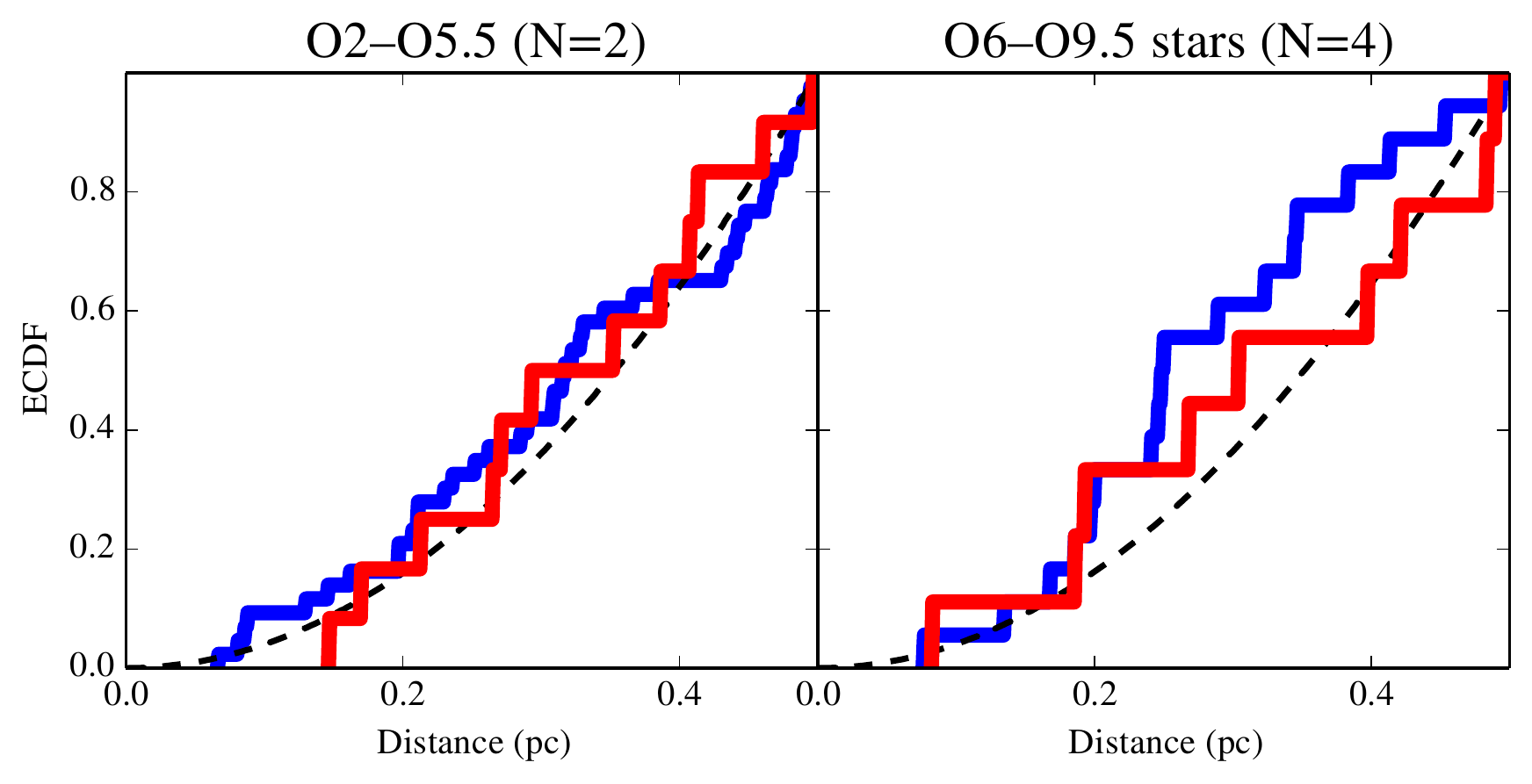}
\caption{Rosette Nebula. \ECDFcaption}
\label{fig:rosette_ECDF}
\end{figure*}

Results of our ECDF analysis for the Rosette Nebula are shown in Figure
\ref{fig:rosette_ECDF}. For both early and late O stars, the distribution of
distances to disk-bearing YSOs very closely follows the distribution predicted
from simple two-dimensional geometry. Given the height of the steps of the
disk-bearing ECDF in both the left and right panels (early and late O stars,
respectively), our data cannot probe the physics of disk destruction over
distances less than $\sim$0.15~pc. Hence, constraints on disk destruction models
derived from these data are weaker than for the other regions discussed in this
subsection, which all show higher densities of disk-bearing YSOs around O stars.
In the remainder of this section, we compare these results with those reported
previously in the literature.

\citet[][hereafter B07]{balogetal2007} test for disk destruction in NGC~2244 by
calculating the distance of each disk-bearing and disk-free YSO to the nearest O
star (of which there are 5 in the region), and compute histograms of these
nearest distances. They report a decrease in disk fraction for YSOs located
within 0.5~pc of an O star.

\begin{figure*}
\centering
\includegraphics[height=\mapheight]{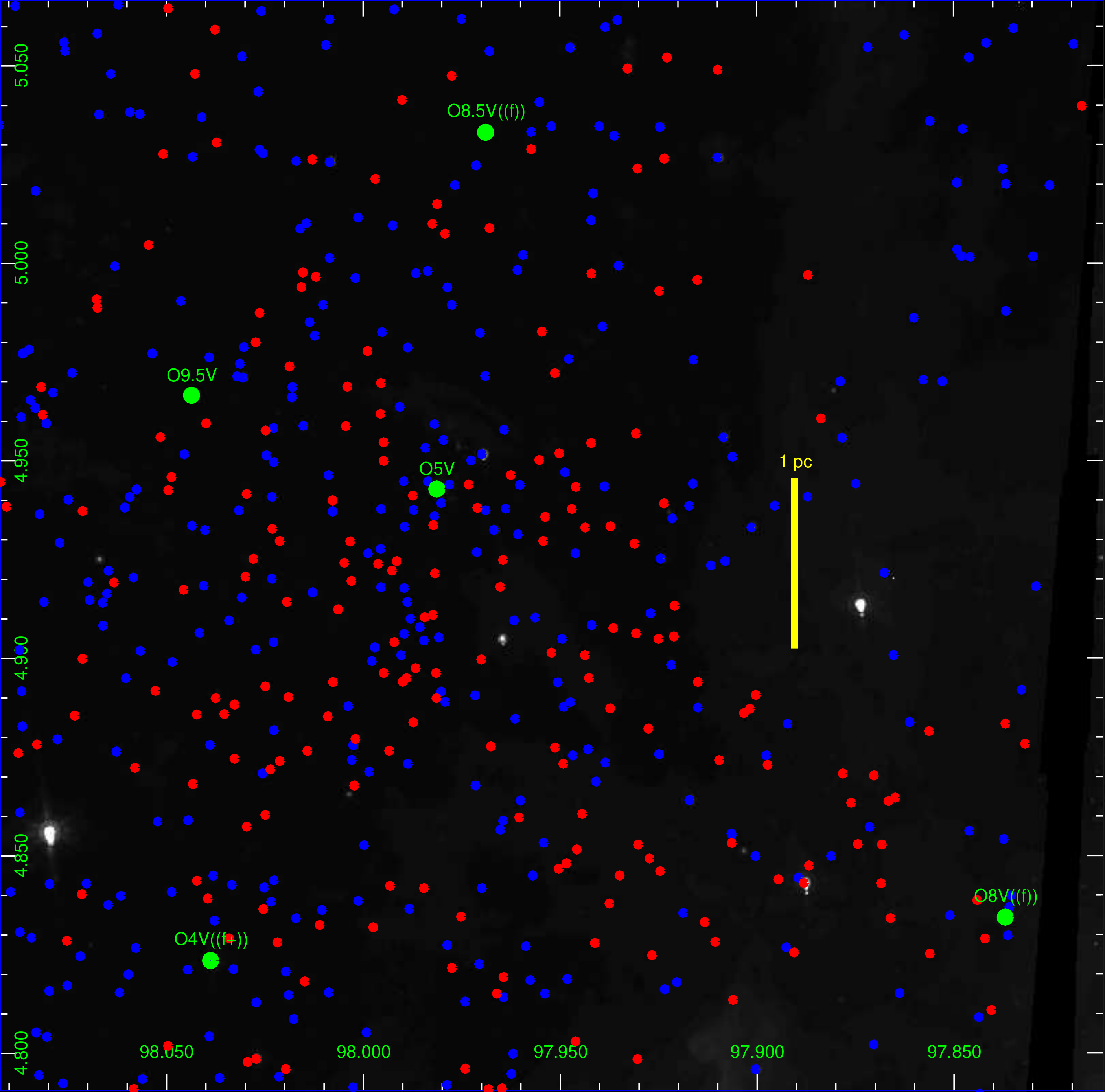}
\caption{RA--Dec map of NGC~2244 (Rosette Nebula), \textbf{using data from
\citet{balogetal2007}} and 8.0~$\micron$ \Spitzer/IRAC image.
\mapcaption{336}{705}{5}{\citet{balogetal2007}} Disk fraction in the cluster is
highly non-uniform, suggesting that the destruction effect reported by
\citet{balogetal2007} is instead the result of disk fraction gradients.}
\label{fig:balog_map}
\end{figure*}

However, the bin widths used in the B07 histograms are much too large to be
sensitive to disk destruction on the scales predicted by theory (the smallest
bin extends to 0.5~pc). Given that the distances under consideration by B07 are
comparable to the size of the cluster, other possible sources of spatially
varying disk fraction must be considered, such as the large cluster age
gradients reported by \citet{getmanetal2014}. Figure~\ref{fig:balog_map} shows a
map of NGC~2244 using the point source data of B07. On visual inspection it is
clear that disk fraction decreases significantly with distance from the cluster
center, even in regions far away from O stars.

\begin{figure*}
\centering
\includegraphics[width=\ECDFwidth]{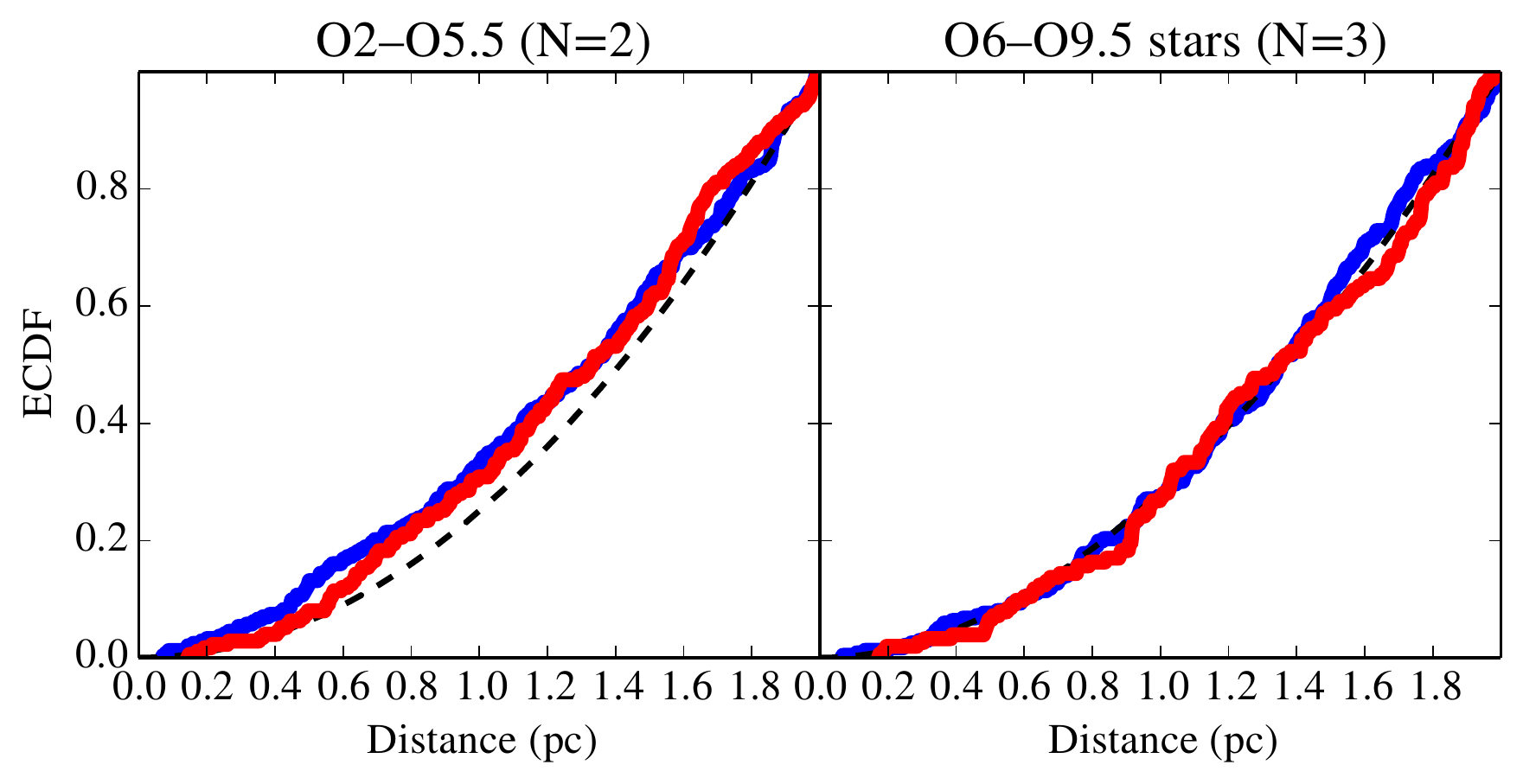}
\caption{NGC~2244 (Rosette Nebula), \textbf{using data from
\citet{balogetal2007}}. \ECDFcaption\ The distributions for the disk-bearing YSOs
do not appear to drop off suddenly at short distances as expected for disk
destruction.}
\label{fig:balog_ECDF}
\end{figure*}

To further explore this problem, in Figure~\ref{fig:balog_ECDF}, we apply the
ECDF analysis used in the current work to the data of B07. This time we extend
our analysis to larger distances due to the large ($\sim$0.5~pc) radius of
effect for disk destruction claimed by B07. We find that the distributions of
disk-bearing YSOs in NGC~2244 increase smoothly, showing no avoidance effect.
The ECDF of disk-bearing YSOs for late O stars (right panel) is somewhat
irregularly shaped, including at $\sim$1~pc and $\sim$1.6~pc. This is most
likely explained by intrinsic disk fraction gradients and missing sources due to
PAH contamination rather than disk destruction.

In the current work, we avoid the potential problems of binning by using ECDFs,
and we restrict our analysis to distances within 0.5~pc of O stars in order to
ignore large-scale gradients in the distribution of disks across the region
unrelated to disk destruction. If disk destruction acts gradually over
$\sim$0.5~pc distances as claimed by B07 (though this is very unlikely), then
the effect would not be distinguishable from other astrophysical and
observational sources of apparent disk fraction gradients. We therefore conclude
that the disk destruction result reported by B07 is not reliable, and our ECDF
analysis of their data appears to refute it.

\subsubsection{NGC~6357}
\label{subsubsect:ngc6357}

\begin{figure*}
\centering
\includegraphics[width=\ECDFwidth]{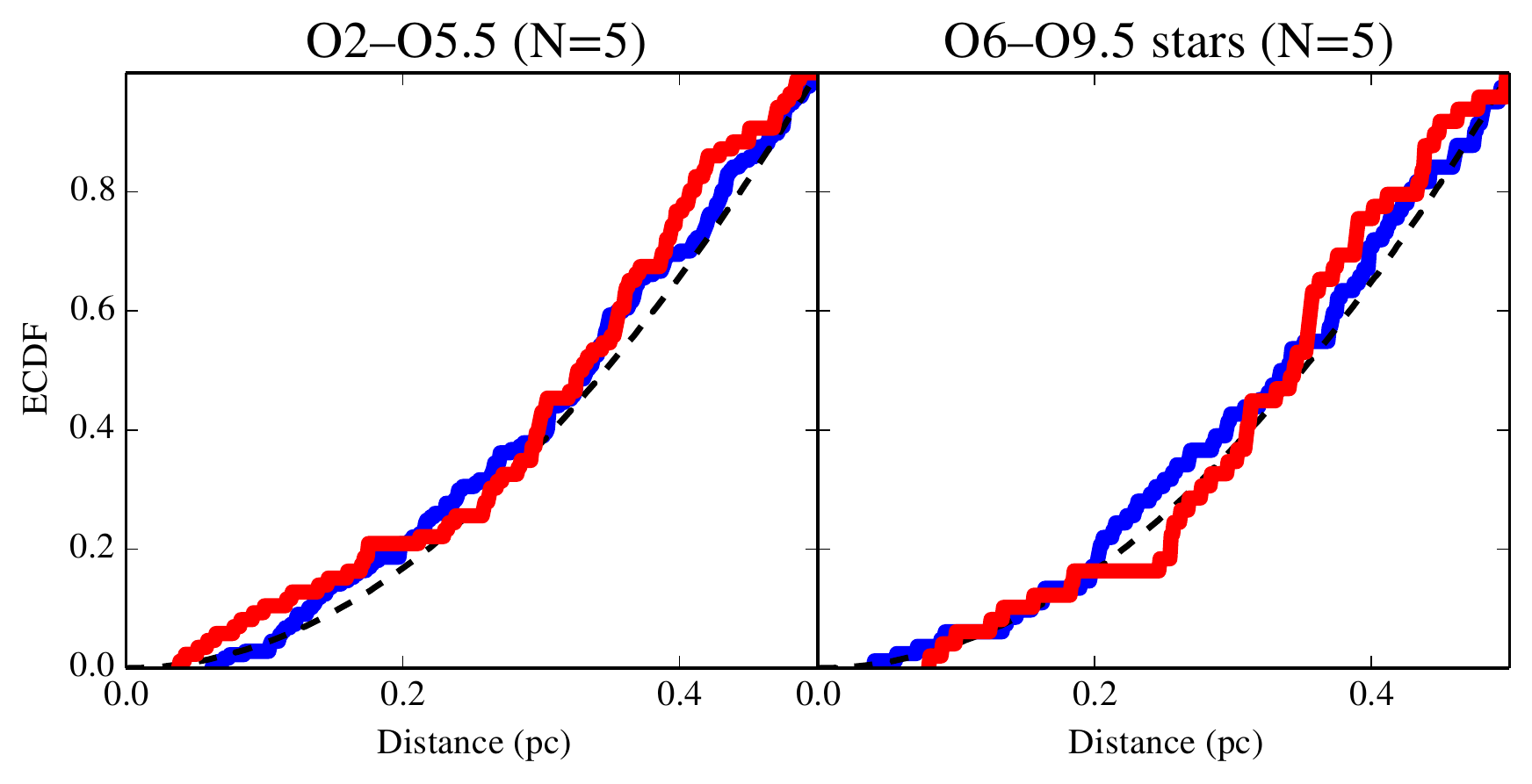}
\caption{NGC~6357. \ECDFcaption}
\label{fig:ngc6357_ECDF} \end{figure*}

Figure~\ref{fig:ngc6357_ECDF} shows results for NGC~6357. Both the disk-bearing
and disk-free distributions closely follow the predicted parabolas. Disk-bearing
YSOs even show a slight overabundance at short distances to early O stars,
possibly reflecting an age gradient in the region, or simply observational
biases.

\subsubsection{Orion Nebula}
\label{subsubsect:orion}

\begin{figure*}
\centering
\includegraphics[width=\ECDFwidth]{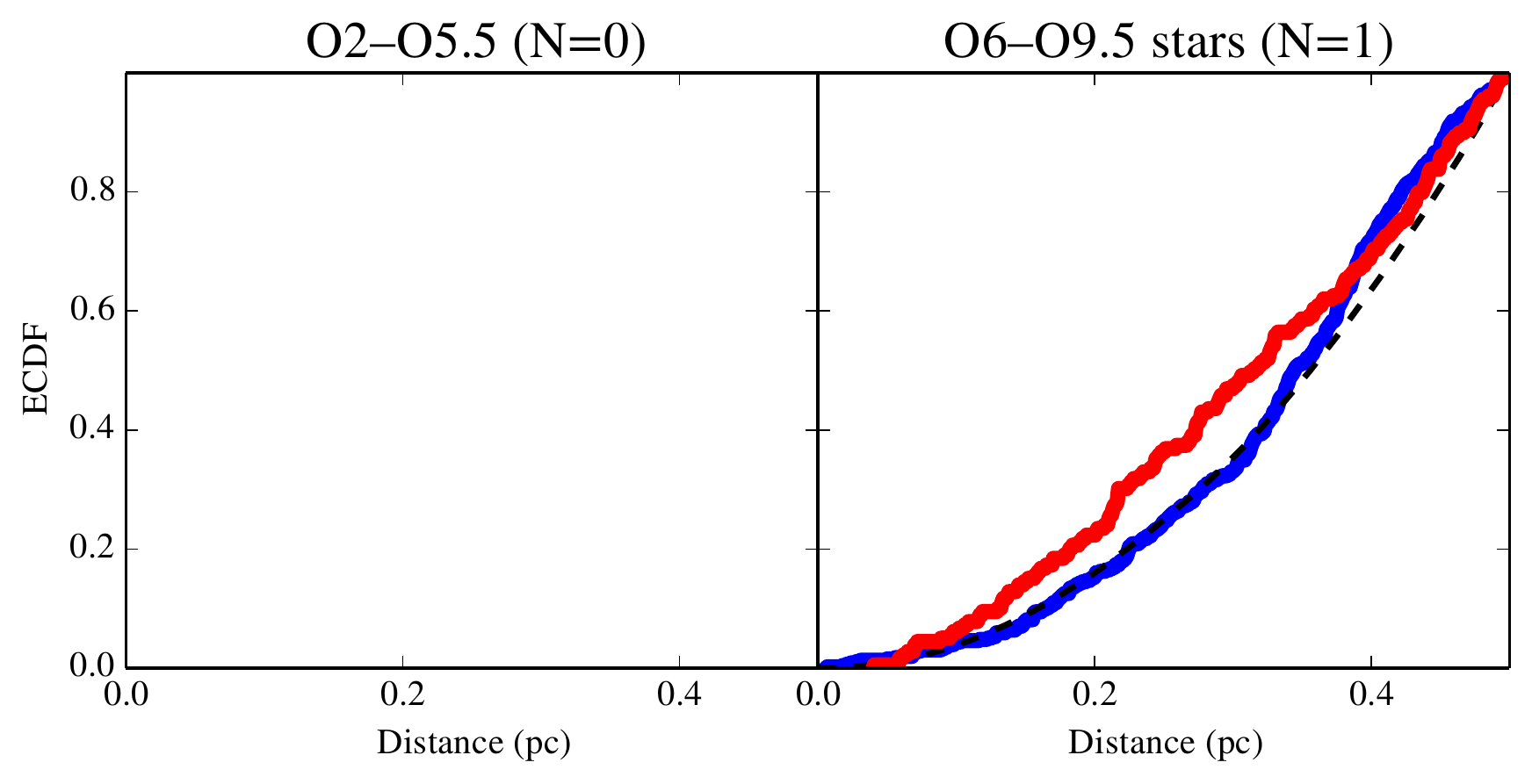}
\caption{Orion Nebula. \ECDFcaption}
\label{fig:orion_ECDF}
\end{figure*}

Figure~\ref{fig:orion_ECDF} shows results for the Orion Nebula. Unlike the
other five regions previously discussed in this subsection, the Orion Nebula
contains no early-type O stars. \tooc\ is omitted from our analysis due to PAH
contamination. Both the disk-bearing and disk-free YSOs show a tendency to
cluster around the one non-PAH excluded late O star (spectral type O9.5) in the
region, as both ECDFs are significantly steeper than the predicted parabolic
CDF at short distances. This is the clearest non-detection of inner disk
destruction (within MYStIX completeness limits) due to the large density of
disk-bearing YSOs around the late O stars (i.e., the very small steps in the
disk-bearing ECDF).

It is interesting to note that the majority of the 178 Orion proplyds
identified by \citet{riccietal2008} with Hubble appear in the MYStIX catalog as
infrared excess (i.e., disk-bearing) sources in the MYStIX sample, with the
notable exception of about a dozen proplyds close to \tooc\ not seen by MYStIX
due to PAH contamination. Hence, many of the disk-bearing YSOs near O stars in
the MYStIX sample are likely in the process of being photoevaporated. The test
for \textit{complete} disk destruction in the current work can at most only
place upper limits on photoevaporative mass-loss rates. Direct estimation of
mass-loss rates requires either spectroscopic analysis of individual disks
\citep[e.g.,][]{henneyodell1999}, or measurements of individual disk masses
\citep[e.g.,][]{mannwilliams2010} combined with age estimates.

\section{Discussion and conclusions}
\label{sect:conclusions}

\subsection{Observational findings}
\label{subsect:obs}

To summarize our results for the 20 MYStIX regions under study,

\begin{itemize}

\item seven regions lack O stars free of PAH contamination
(\S~\ref{subsect:cat1}),

\item seven regions show too few disk-bearing YSOs around O stars for disk
destruction to be detected (\S~\ref{subsect:cat2}),

\item four regions show clear evidence that inner disk depletion has
\textit{not} occurred within $\sim$0.2~pc of early O stars (Eagle, NGC~6357,
Orion, and Rosette; \S~\ref{subsect:cat3}), and

\item two regions show inconclusive results (Carina and M~17;
\S~\ref{subsect:cat3}).

\end{itemize}

Our study of protoplanetary disk destruction by external photoevaporation has
several advantages over past studies, but also several limitations. In its
favor, our sample covers 20 massive star-forming regions out to $\sim$3~kpc
distance, allowing us to examine disks around a large sample of O stars,
including the hottest and most luminous O2--O5 stars where disk destruction
should be most effective. The MYStIX survey provides an unusually large sample
of both high and low mass stars so that even a slight but consistent disk
destruction effect in a given region could be discerned when averaged over many
O stars (our data, which provide only a binary determination of disk-bearing
versus disk-free, cannot reliably detect disk destruction around individual O
stars due to the relative sparseness of surrounding disk-bearing YSOs).

The size of the MYStIX dataset notwithstanding, our search for spatial
avoidance of O stars by disks faces numerous difficulties: non-detection of
disks due to PAH contamination; incompleteness of the disk sample;
non-detection of disks due to crowding; and astrophysical gradients in the
spatial distribution of disks due to clustering behavior, age gradients, etc.

The problem of crowding in dense, distant regions may be the most significant
difficulty in detecting disk destruction with MYStIX, as it is precisely these
densely-populated regions where disk destruction is most likely to operate (due
to small interstellar distances). In the case of the MYStIX sample, Trumpler~14
in the Carina Nebula and NGC~6618 in M~17 are two of the most densely-populated
regions with some of the earliest O stars, and they are both located far away
($\sim$2~kpc), making their members even more difficult to resolve. It is for
the aforementioned reasons that, using MYStIX data as well as the data of
\citet{guarcelloetal2009} and \citet{balogetal2007}, we find that the apparent
distribution of disk-bearing YSOs can vary significantly across a given region
on spatial scales similar to those associated with external photoevaporation by
massive stars. All of these effects are in principle capable of giving rise
both to false positive and false negative detections of disk destruction.

There are several ways in which the use of point source data in
identifying spatial avoidance of massive stars by disks could be improved in
future works. While nebular contamination will always make disk detection
difficult, the problem of stellar crowding, due in part to large region
distances, could be alleviated with high-resolution (PSF$\ll$1") infrared
photometry of these dense cluster cores using 8~m-class telescopes with adaptive
optics, or with the space-based James Webb Space Telescope. The dense OB
association in Trumpler~14 (Carina Nebula) is highly crowded but has relatively
little nebulosity, and would therefore make an excellent target for
high-resolution observations. Deeper observations would also be useful for
detect disks around lower mass stars, particularly at high resolutions given the
large numbers of low mass stars in cores of dense, massive star-forming regions.
Such observations would be particularly useful for examining disk destruction in
some of the regions discussed in Section~3.2, with low apparent surface
densities of disk-bearing YSOs which may in many cases be the result of
observational limits rather than intrinsically low number densities.

Another way of improving spatial avoidance-based detection of disk
destruction would be the incorporation of far-infrared observations. Photometry
from \Spitzer/MIPS and \textit{Herschel}, for instance, could probe disks at
greater radii (tens of AU), though the relatively low spatial resolution of
these instruments would make disk detection difficult in dense, distant
regions. Nonetheless, far-infrared observations could help to resolve questions
of how far photoevaporation proceeds into a given disk, which potentially has
important implications for the evolution of planetary systems (discussed
further in the following section).

\subsection{Astrophysical considerations}
\label{subsect:phys}

Our non-detection of disk destruction around dozens of O stars in several
densely-populated massive star-forming regions is consistent with Orion Nebula
studies that suggest that external photoevaporation by massive stars does not
significantly suppress Solar System-scale planet formation
\citep{eisnercarpenter2006, mannwilliams2009, mannetal2014}.
\citet{mannwilliams2009}, for instance, find that the fraction of disks with
masses comparable to the Minimum Mass Solar Nebula in the Trapezium Cluster is
similar to the fraction found in regions not dominated by O stars. Moreover,
several theoretical works predict that external photoevaporation, as it
disperses the disk starting at the outer edge and working its way inward, ceases
to significantly operate at radii of a few tens of AU from the disk's host star
where accretion timescales may still be relatively large \citep{clarke2007,
adamsetal2004, adamsetal2006, fatuzzoadams2008, mannwilliams2009}. This
particular prediction is supported by findings based on near-/mid-infrared
excesses, namely, the current work and \citet{oliveiraetal2005}, who find that
the disk fraction in NGC~6611 (Eagle Nebula) is consistent with that of regions
of similar age but without O stars.

\citet{andersonetal2013} explore disk models which combine external FUV
photoevaporation and viscous accretion. For a range of Shakura--Sunyaev $\alpha$
viscosities \citep{shakurasunyaev1973} and local FUV field strengths comparable
to or greater than those of the Orion proplyds, they predict disk lifetimes
between $5\times 10^5$~yr and $10^7$~yr, which are not significantly different
from disk lifetimes based on viscous accretion alone. Their predictions based on
nominal values of disk $\alpha$ and FUV field strength are roughly consistent
with the results of the current work, however stronger constraints on
photoevaporative mass loss rates would require the study of somewhat older
regions (as our results only rule out the largest values of disk viscosity and
FUV field strength). The Carina Nebula, whose members range from $\sim$1--4~Myr
in age and which contains a significant number of both early and late O stars
(as well as dense clusters of O stars such as in Trumpler~14), is therefore a
prime target for further studies of disk destruction.

While the effect of external photoevaporation on the inner disk appears limited,
it is possible that external photoevaporation suppresses the formation of
planetesimals, including comets and the seeds of massive planets, in the outer
disk \citep[beyond 10--30~AU from the disk host star;][] {adamsetal2004,
adamsetal2006, fatuzzoadams2008}, or halts the inward migration of Jovian
planets formed in the outer disk (or even contributes to outward migration due
to positive torques from gas within the planet's orbit). \citet{adamsetal2004}
suggest that external photoevaporation of gas in the outer disk in the early
Solar System may explain the relative gas poorness of Uranus and Neptune,
although other explanations have been offered \citet[e.g.,][]{thommesetal2002}.

The observed ubiquity of exoplanets \citep{boruckietal2011,
petiguraetal2013, tuomietal2014} clearly suggests that the overall effect of
external photoevaporation on planet formation must be limited, either because
complete disk destruction by massive stars is rare, or because planets are
generally able to form before disk destruction occurs \citep[the findings of
][for instance, suggest that planet formation during protostellar collapse is
required to explain the observed population of exoplanets]{najitakenyon2014}.
Nonetheless, understanding the effect of external photoevaporation on planet
formation and migration in detail will require stronger constraints on the
timescales of external photoevaporation as well as of planet formation and
migration. In terms of the former, while the methods used in the current work
provide loose constraints on photoevaporation timescales, other more direct
methods for constraining photoevaporative mass loss rates are available.
\citet{henneyodell1999}, for instance, use high-resolution spectroscopy to
measure proplyd sizes and outflow velocities. Direct measurements of disk
masses using sub-mm observations \citep{mannwilliams2009b, mannwilliams2010,
mannetal2015}, can be in principle be coupled with age estimates to calculate
mass loss rates as a function of projected distance from O stars. Mass loss
rates are relatively constant over a disk's lifetime \citep{andersonetal2013},
therefore disk destruction timescales can be calculated as the ratio of assumed
initial disk masses to observed mass loss rates. Many infrared-excess objects
in the MYStIX sample, particularly near early O stars in the regions discussed
in Section~3.3, are good candidates for such observations.

\acknowledgements

We appreciate insightful discussions with Matthew Povich (California State
Polytechnic University at Pomona). We thank P. Broos (Penn State), L. Townsley
(Penn State), T. Naylor (University of Exeter), M. Povich (California State
Polytechnic University), and K. Luhman (Penn State) for development of X-ray
and IR analysis tools, and for participation in production of MYStIX catalogs.
The MYStIX project has been supported at Penn State by NASA grant NNX09AC74G,
NSF grant AST-0908038, and the \Chandra/ACIS Team contract SV4-74018 issued by
SAO/CXC under contract NAS8-03060.

\newpage

\appendix
\section{MYStIX region maps} \label{sect:app1}
\begin{figure*}[h]
\centering
\includegraphics[width=\mapwidth]{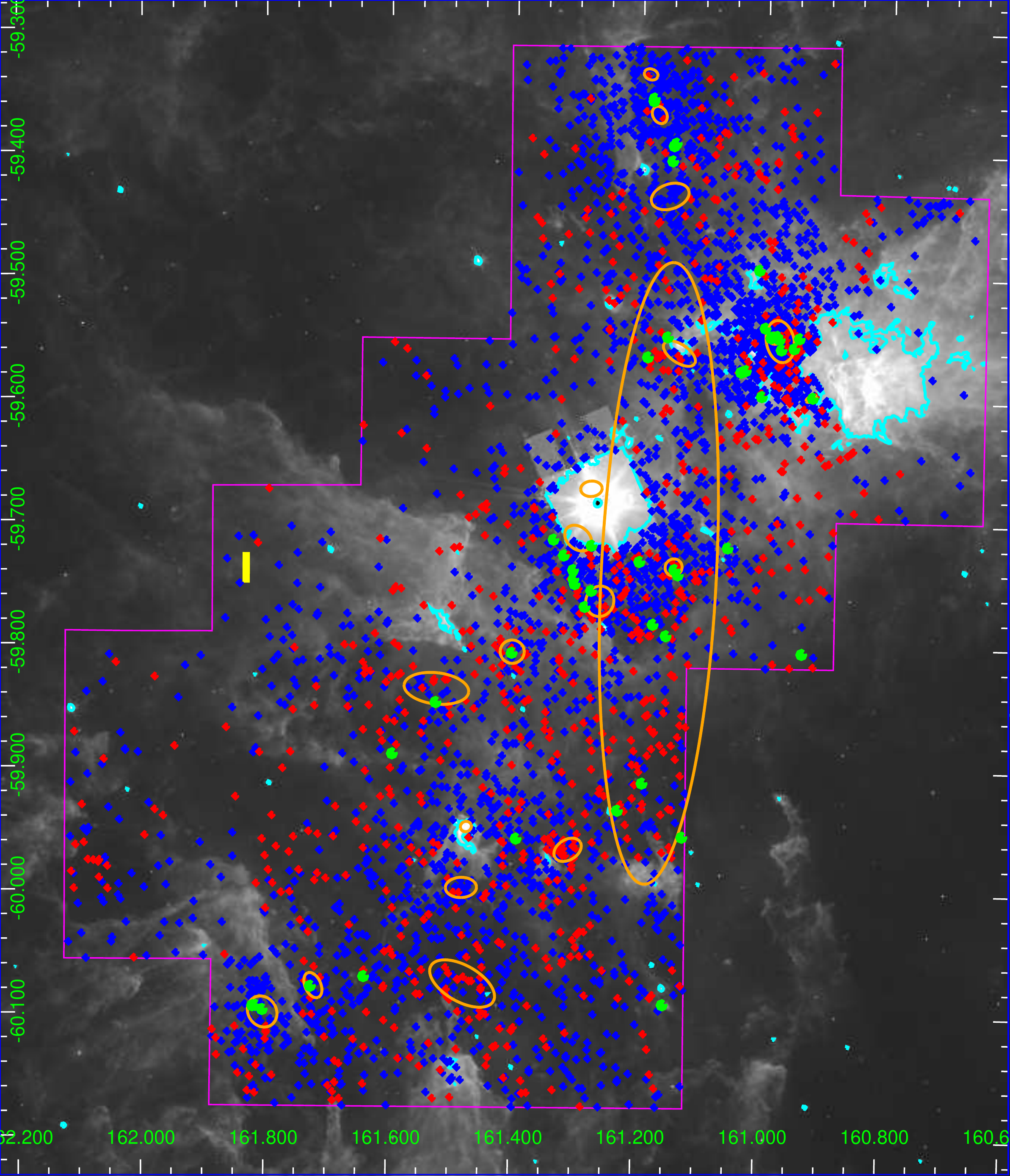}
\caption{Carina Nebula. \appendixcaption}
\label{fig:carina_map}
\end{figure*}

\begin{figure*}[h]
\centering
\includegraphics[width=\mapwidth]{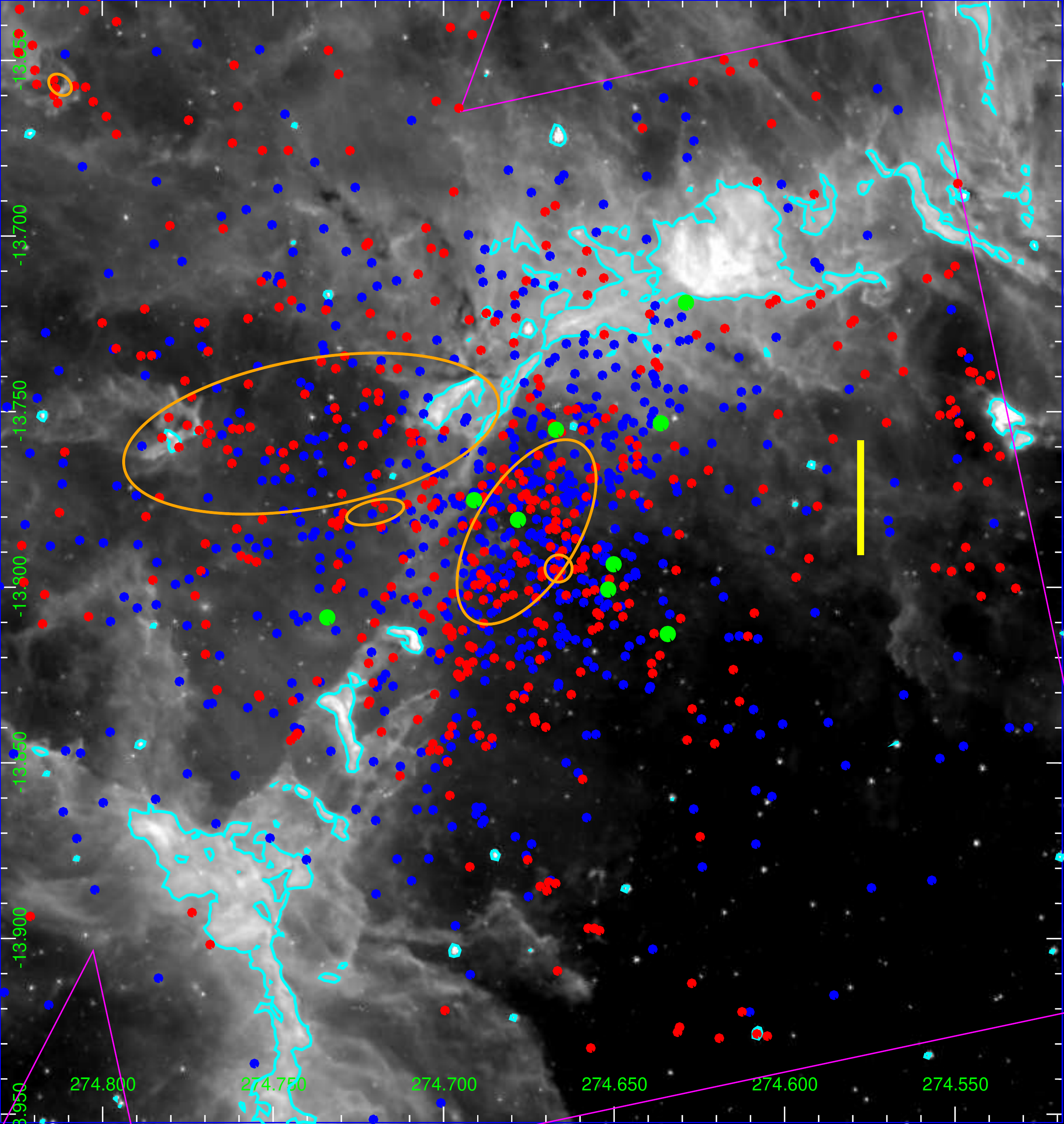}
\caption{Eagle Nebula. \appendixcaption}
\end{figure*}

\begin{figure*}[h]
\centering
\includegraphics[width=\mapwidth]{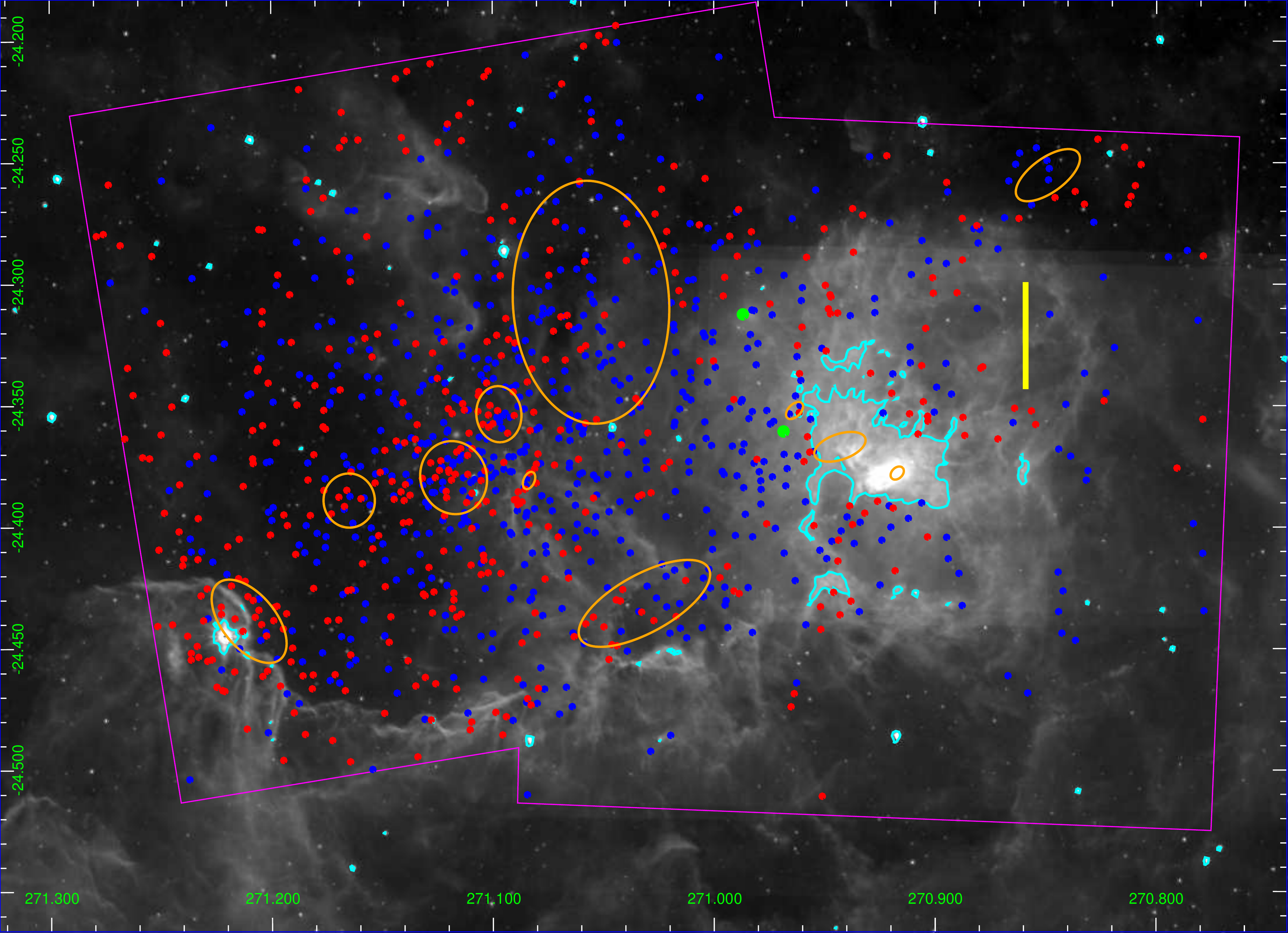}
\caption{Lagoon Nebula. \appendixcaption}
\end{figure*}

\begin{figure*}[h]
\centering
\includegraphics[width=\mapwidth]{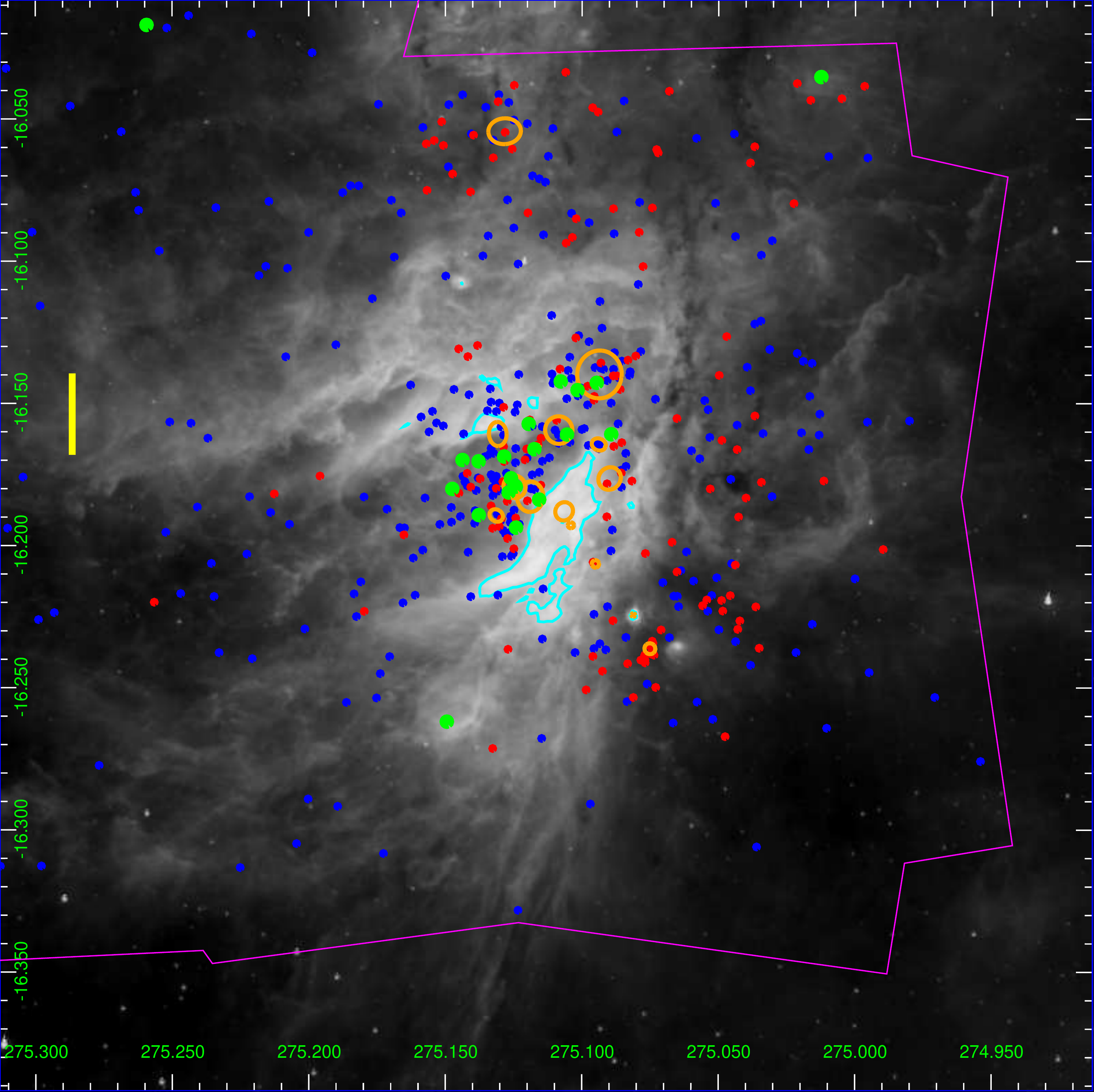}
\caption{M~17. \appendixcaption}
\end{figure*}

\begin{figure*}[h]
\centering
\includegraphics[width=\mapwidth]{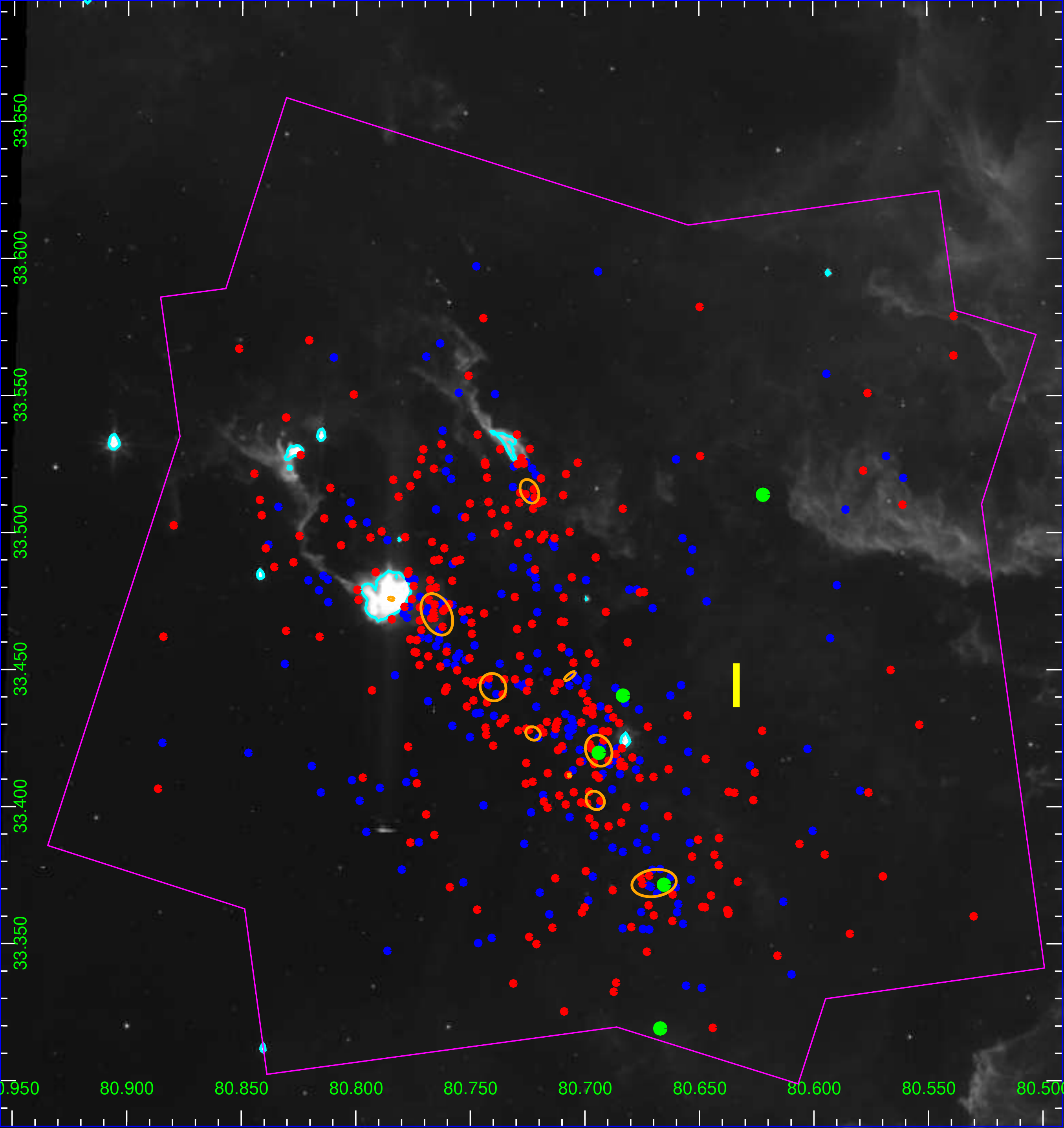}
\caption{NGC~1893. \appendixcaption}
\end{figure*}

\begin{figure*}[h]
\centering
\includegraphics[width=\mapwidth]{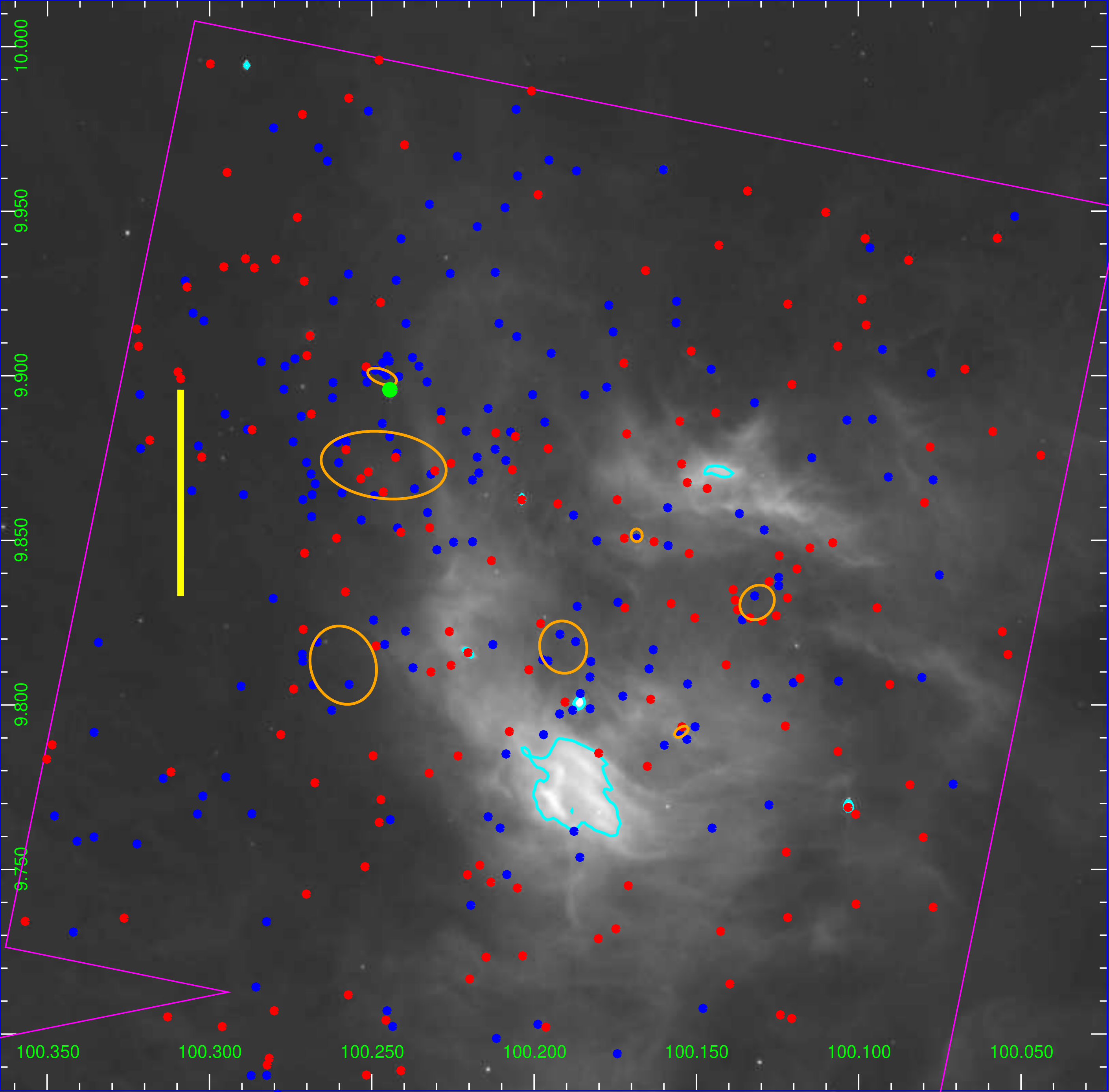}
\caption{NGC~2264. \appendixcaption}
\end{figure*}

\begin{figure*}[h]
\centering
\includegraphics[width=\mapwidth]{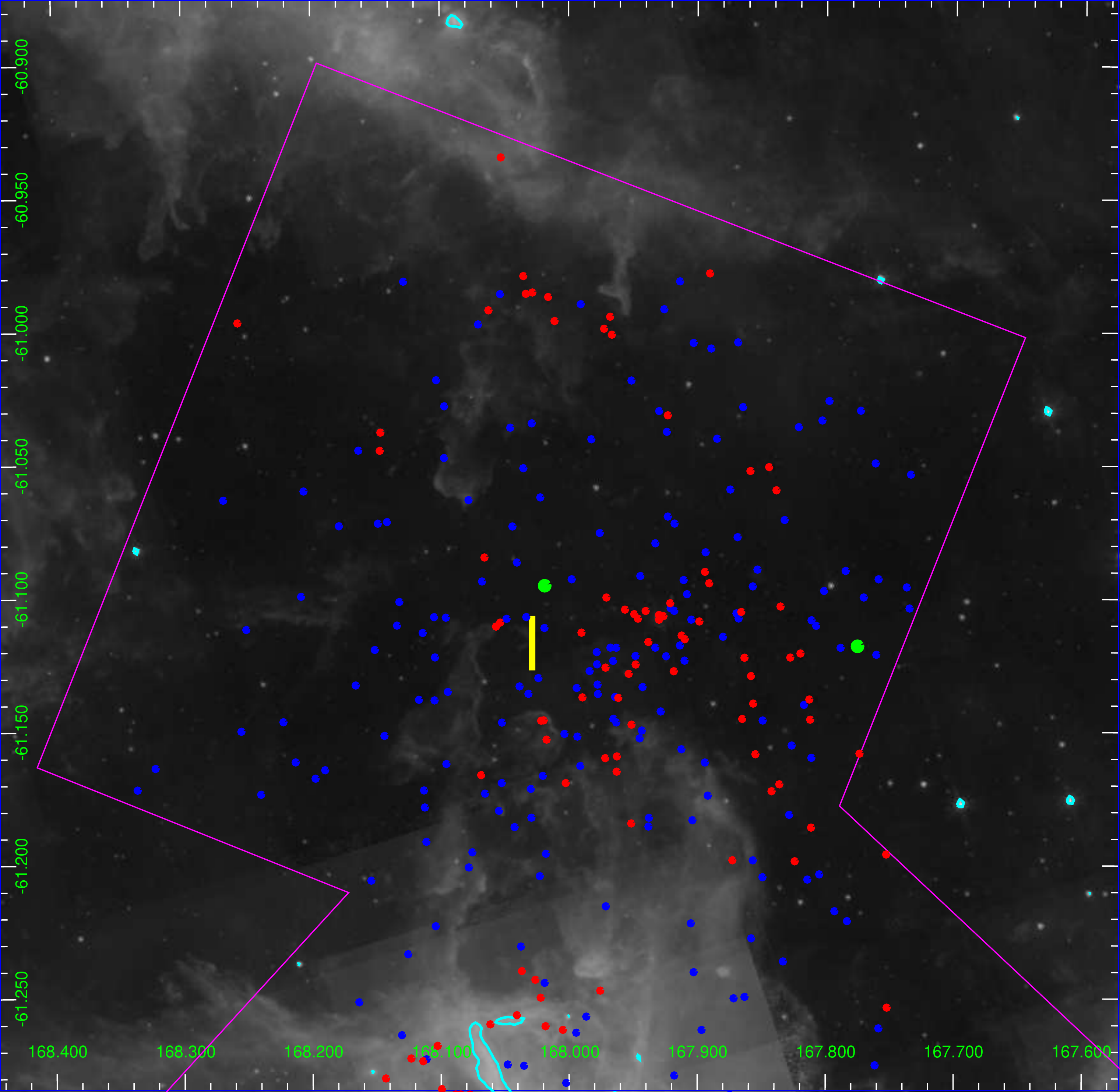}
\caption{NGC~3576. \appendixcaption}
\end{figure*}

\begin{figure*}[h]
\centering
\includegraphics[width=\mapwidth]{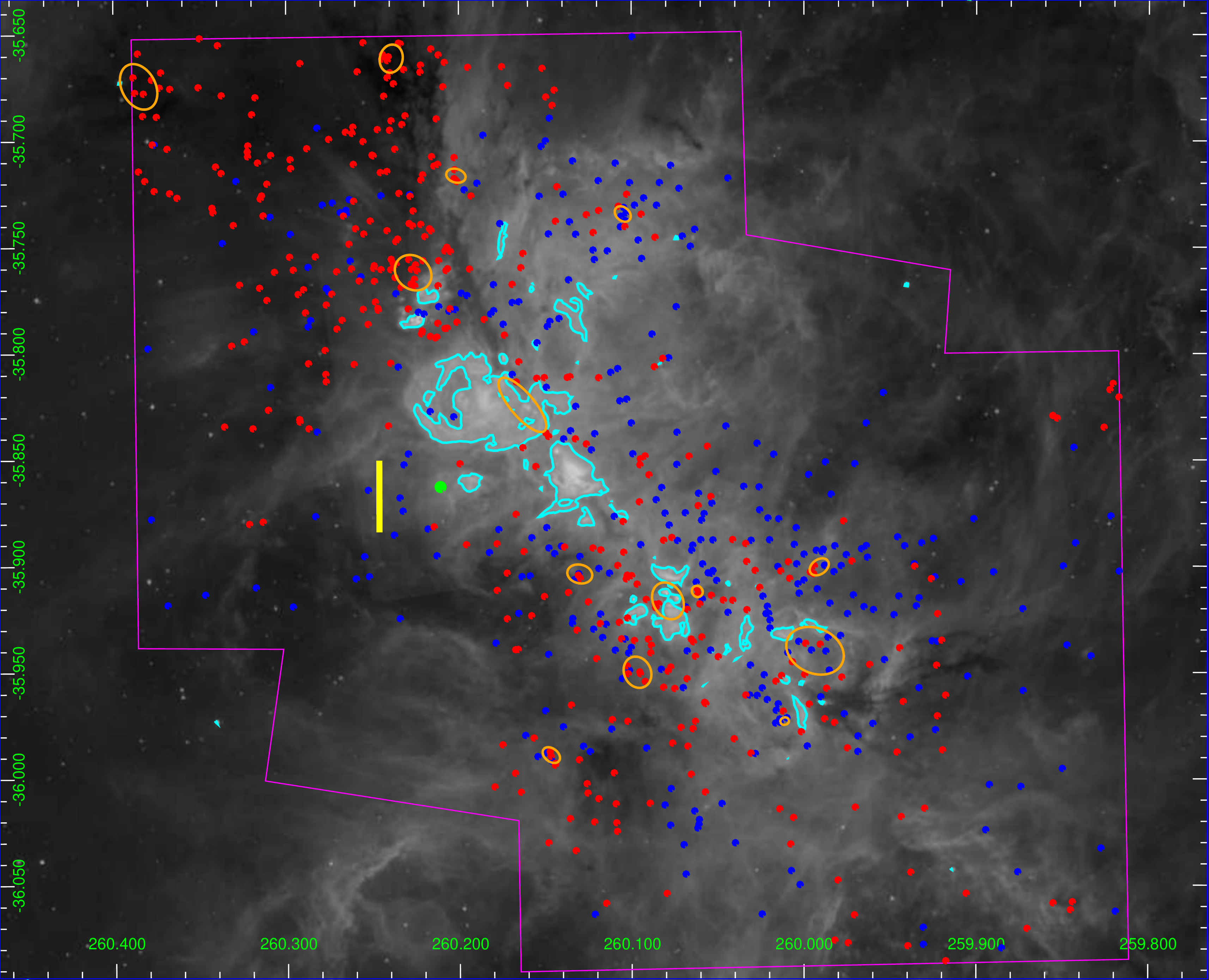}
\caption{NGC~6334. \appendixcaption}
\end{figure*}

\begin{figure*}[h]
\centering
\includegraphics[width=\mapwidth]{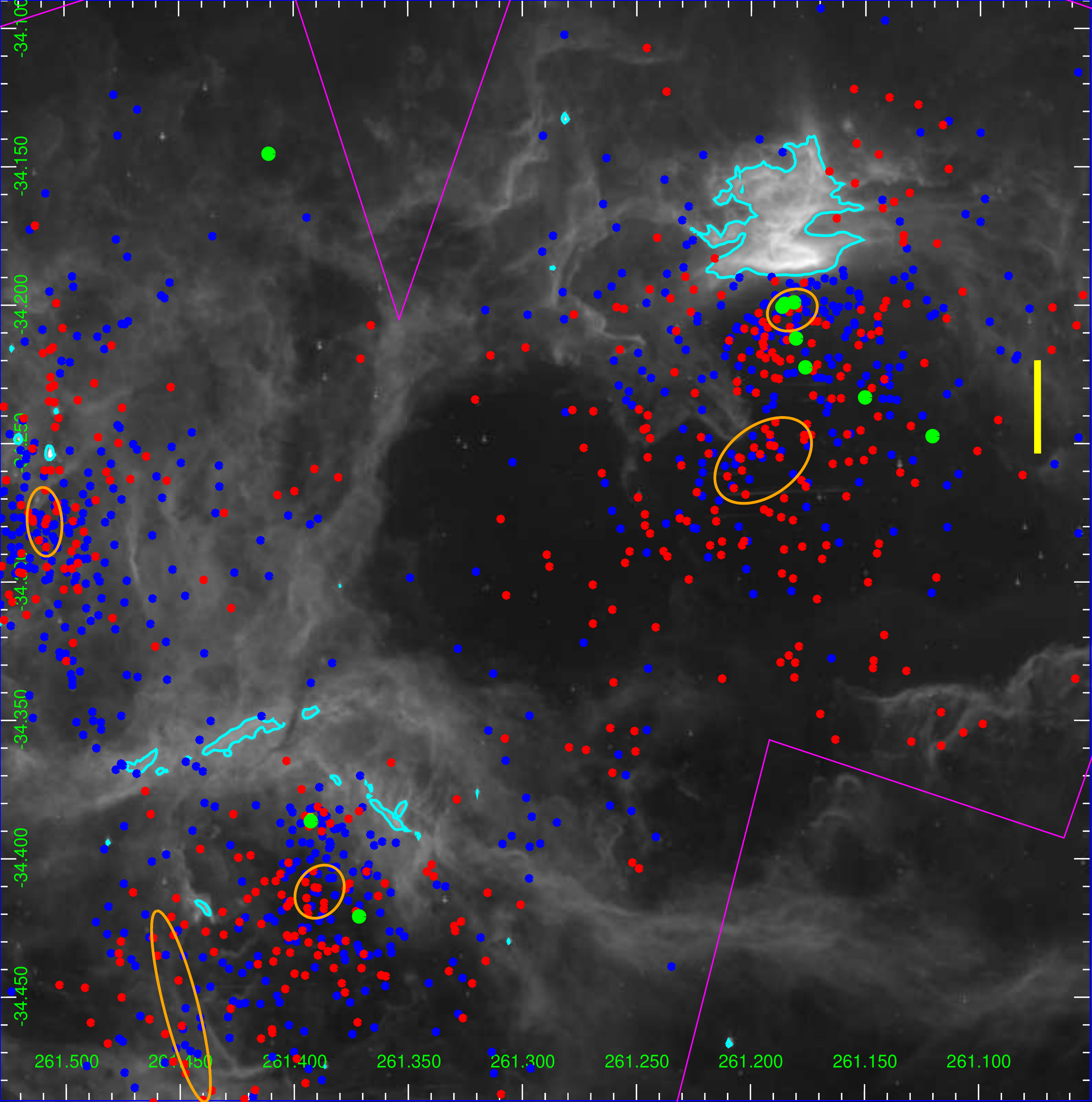}
\caption{NGC~6357. \appendixcaption}
\end{figure*}

\begin{figure*}[h]
\centering
\includegraphics[width=\mapwidth]{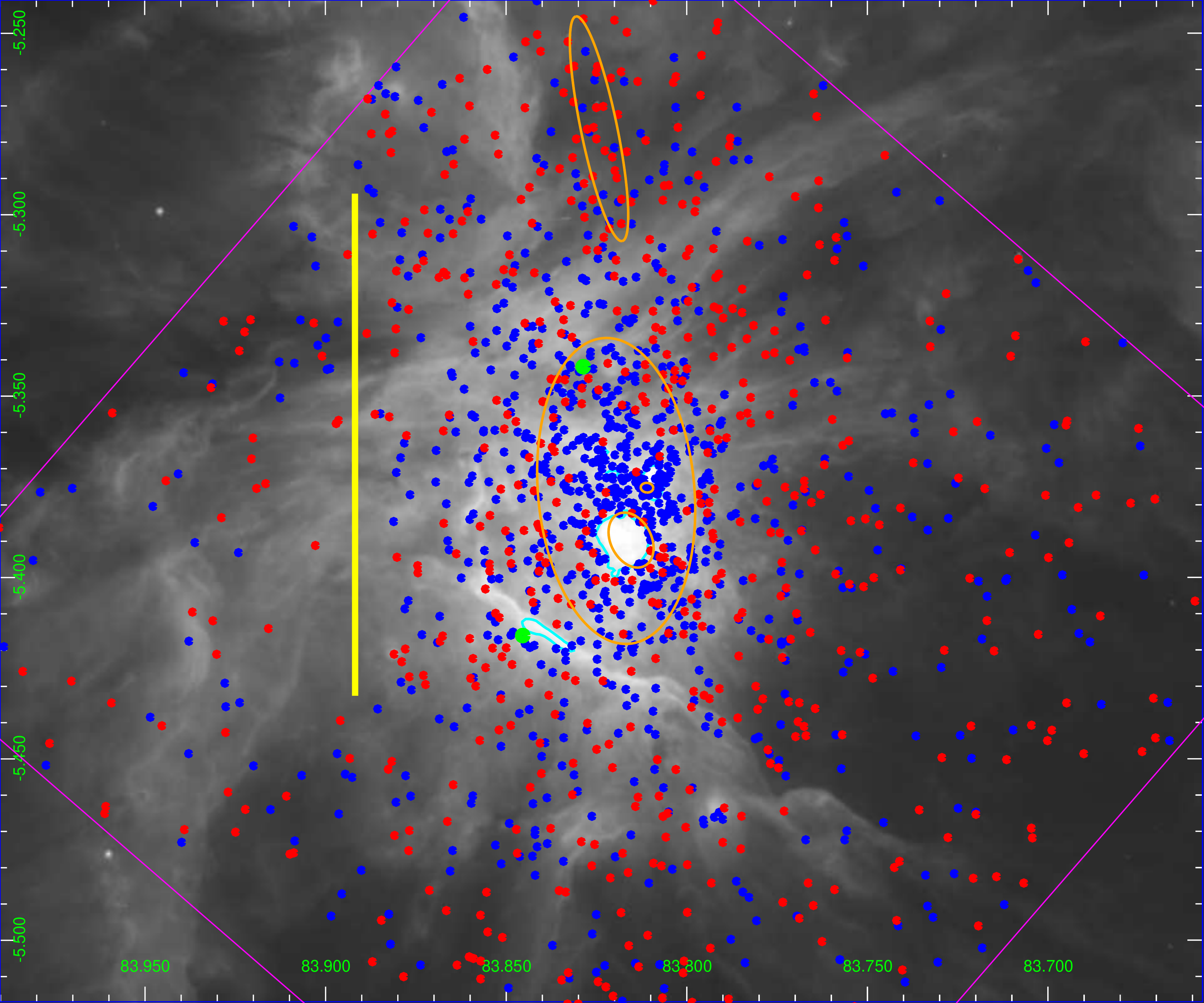}
\caption{Orion Nebula. \appendixcaption}
\end{figure*}

\begin{figure*}[h]
\centering
\includegraphics[width=\mapwidth]{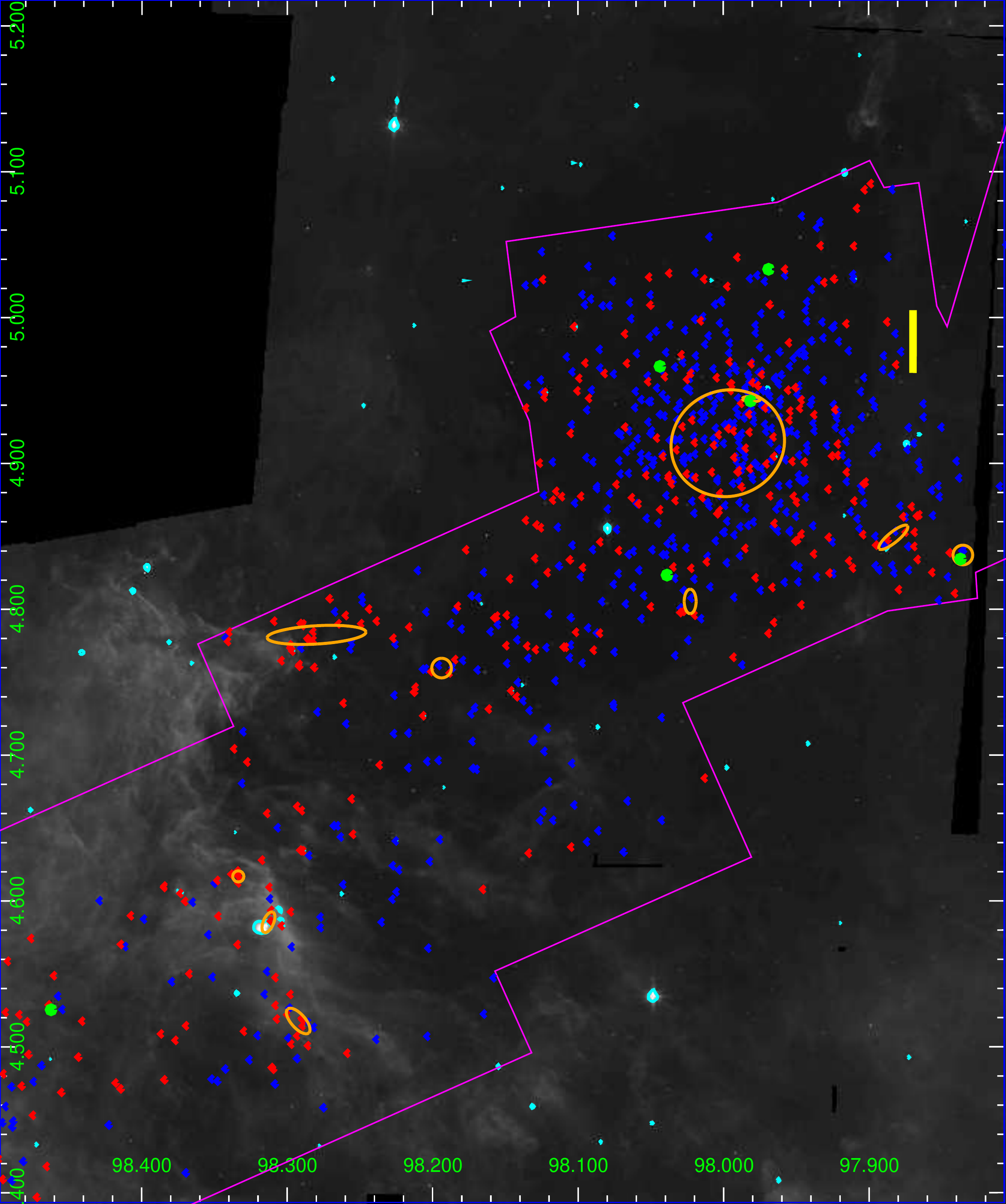}
\caption{Rosette Nebula. \appendixcaption}
\end{figure*}

\begin{figure*}[h]
\centering
\includegraphics[width=\mapwidth]{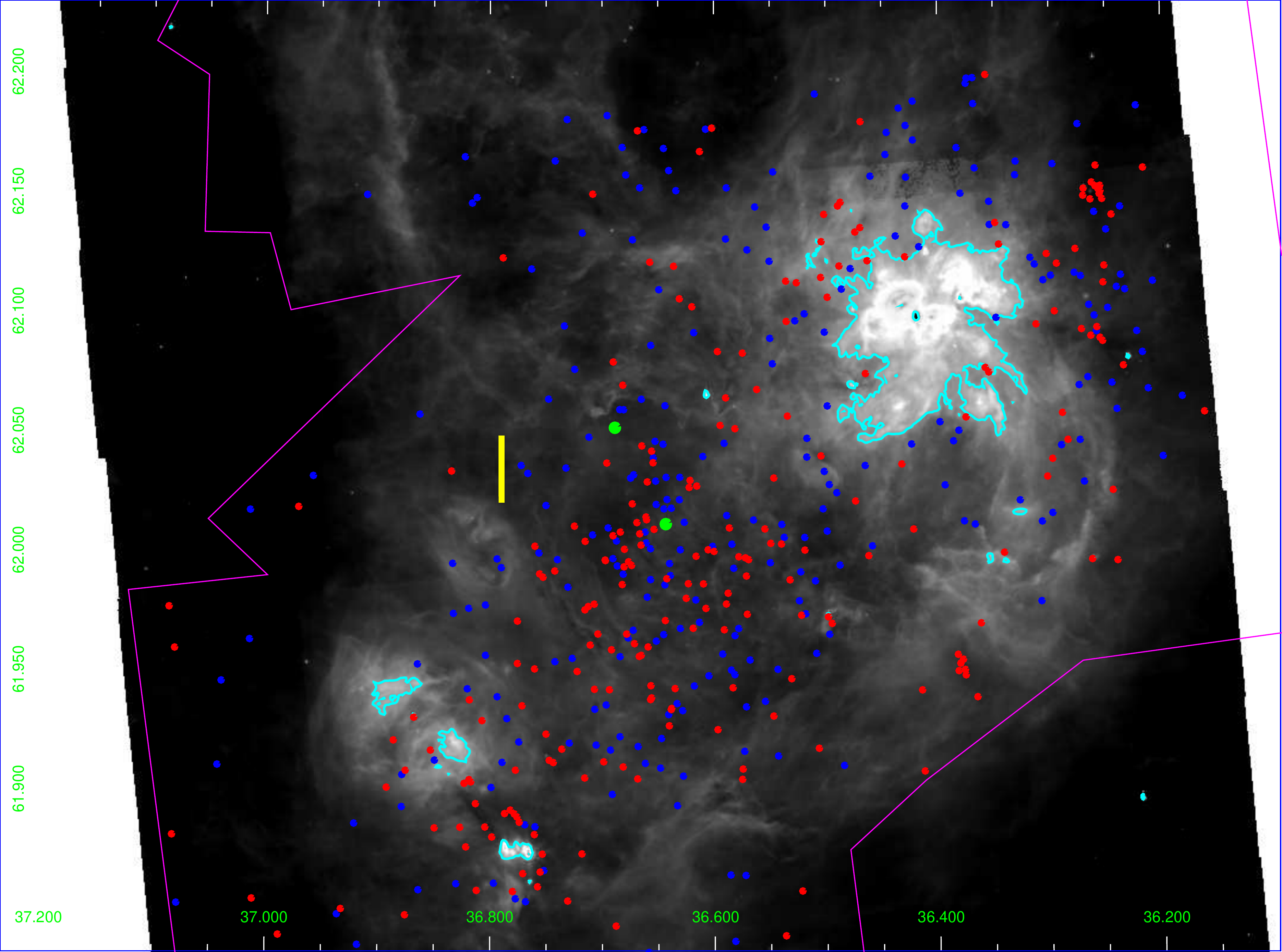}
\caption{W~3. \appendixcaption}
\end{figure*}

\begin{figure*}[h]
\centering
\includegraphics[width=\mapwidth]{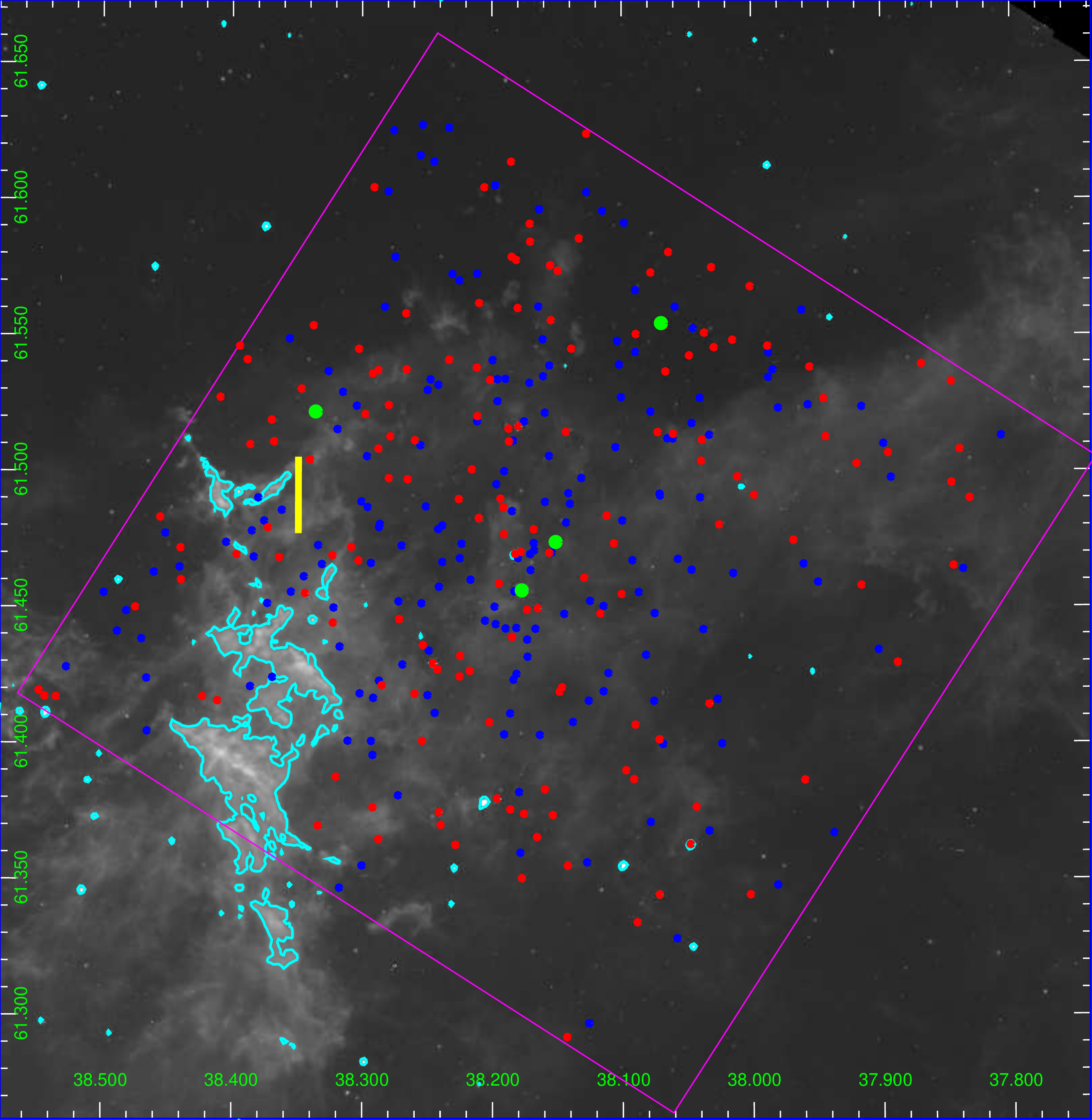}
\caption{W~4. \appendixcaption}
\end{figure*}

\clearpage

\newpage
\section{ECDF results for low-surface density MYStIX regions}\label{sect:app2}

\begin{figure*}[h]
\centering
\includegraphics[width=\ECDFwidth]{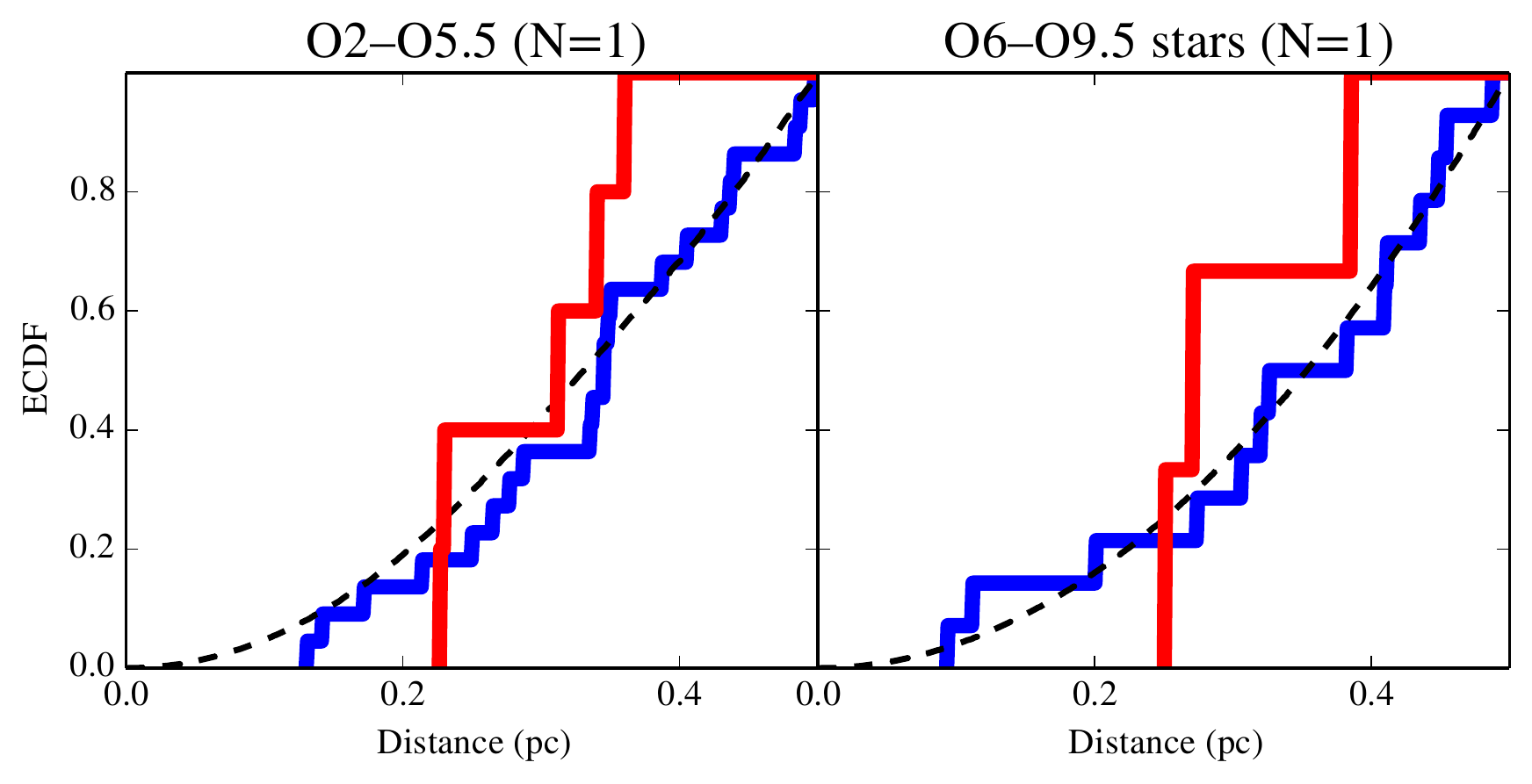}
\caption{Lagoon Nebula. \ECDFcaption\ \ECDFcattwocaption}
\label{fig:app2A}
\end{figure*}

\begin{figure*}[h]
\centering
\includegraphics[width=\ECDFwidth]{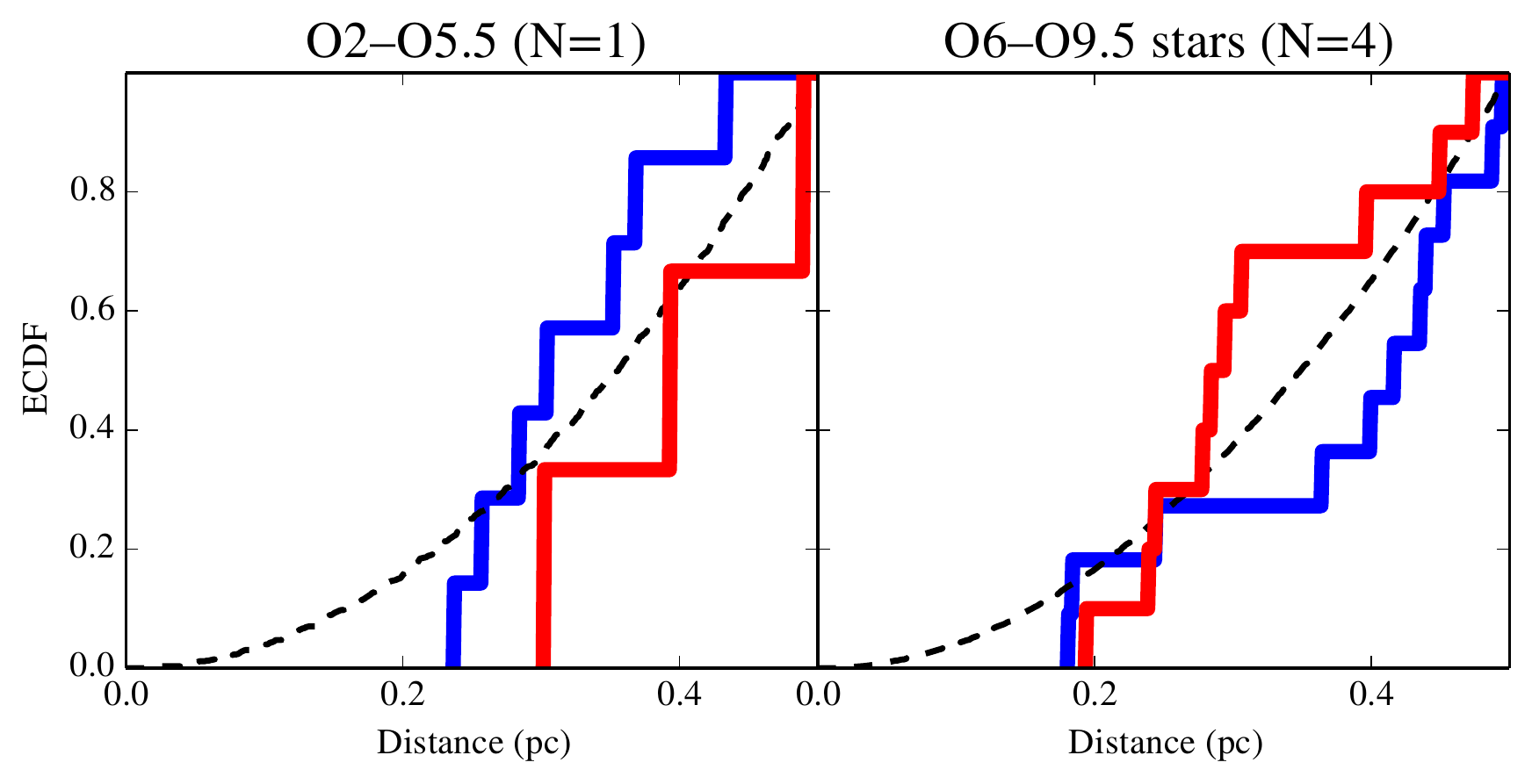}
\caption{NGC~1893. \ECDFcaption\ \ECDFcattwocaption}
\end{figure*}

\begin{figure*}[h]
\centering
\includegraphics[width=\ECDFwidth]{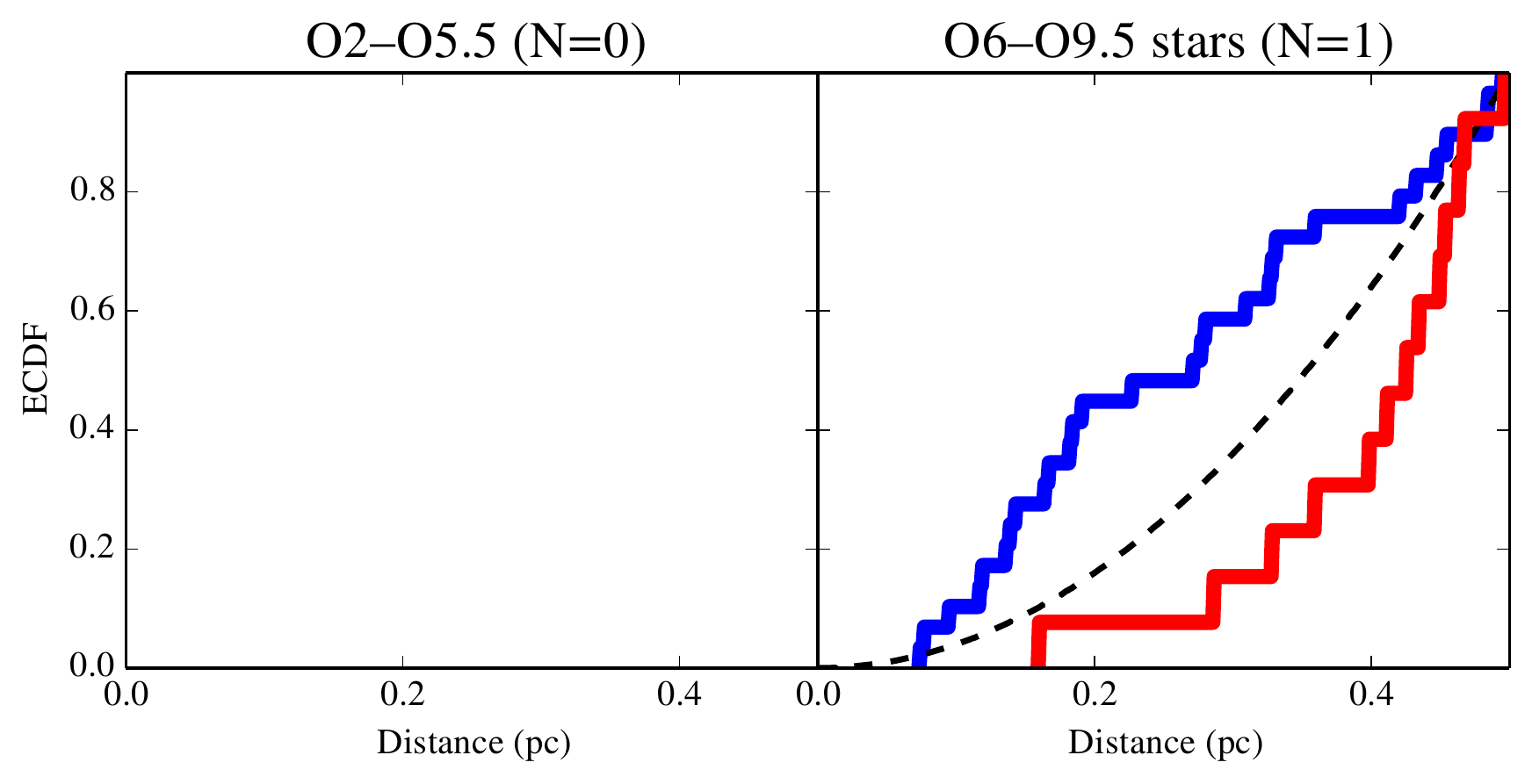}
\caption{NGC~2264. \ECDFcaption\ \ECDFcattwocaption}
\end{figure*}

\begin{figure*}[h]
\centering
\includegraphics[width=\ECDFwidth]{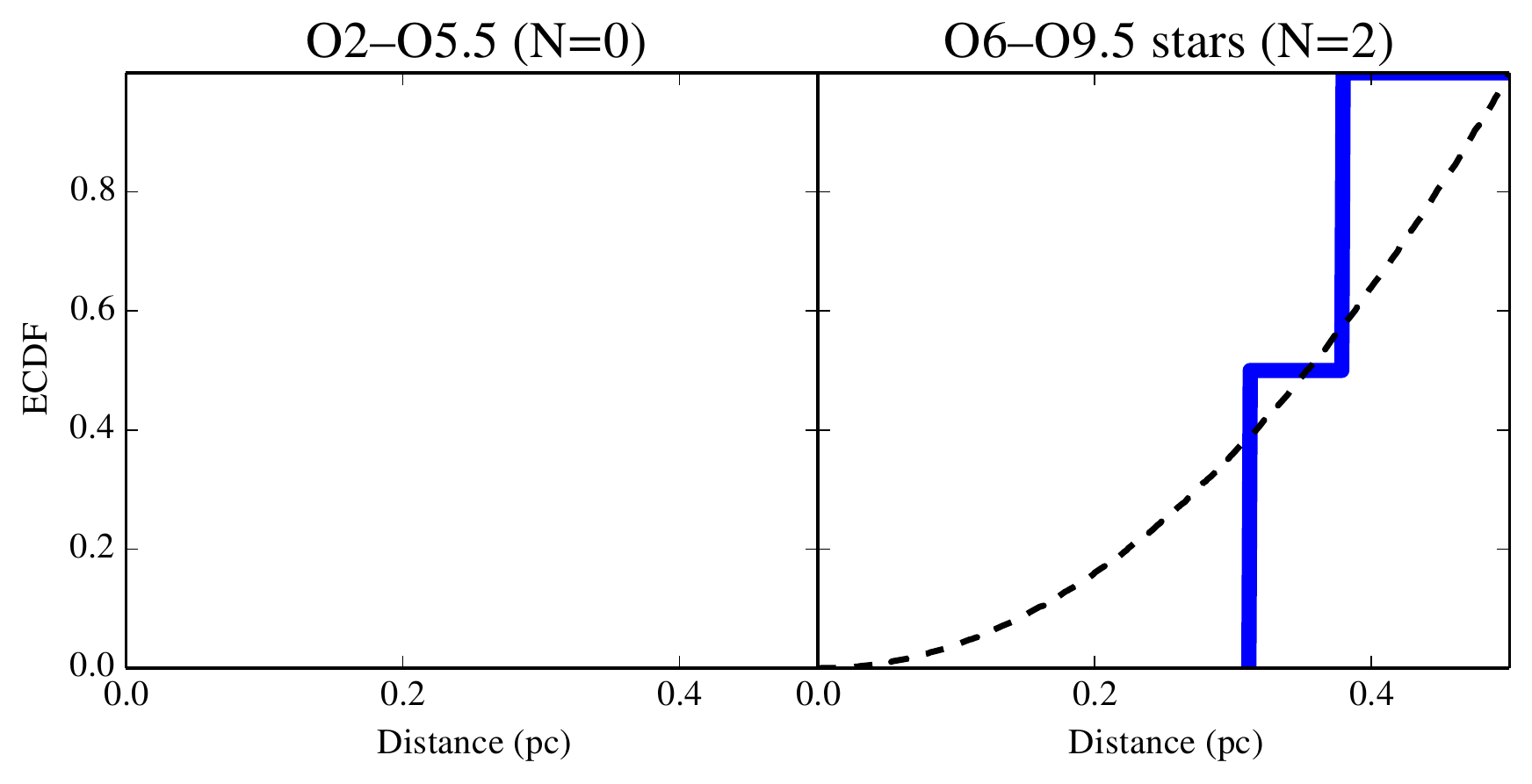}
\caption{NGC~3576. \ECDFcaption\ \ECDFcattwocaption}
\end{figure*}

\begin{figure*}[h]
\centering
\includegraphics[width=\ECDFwidth]{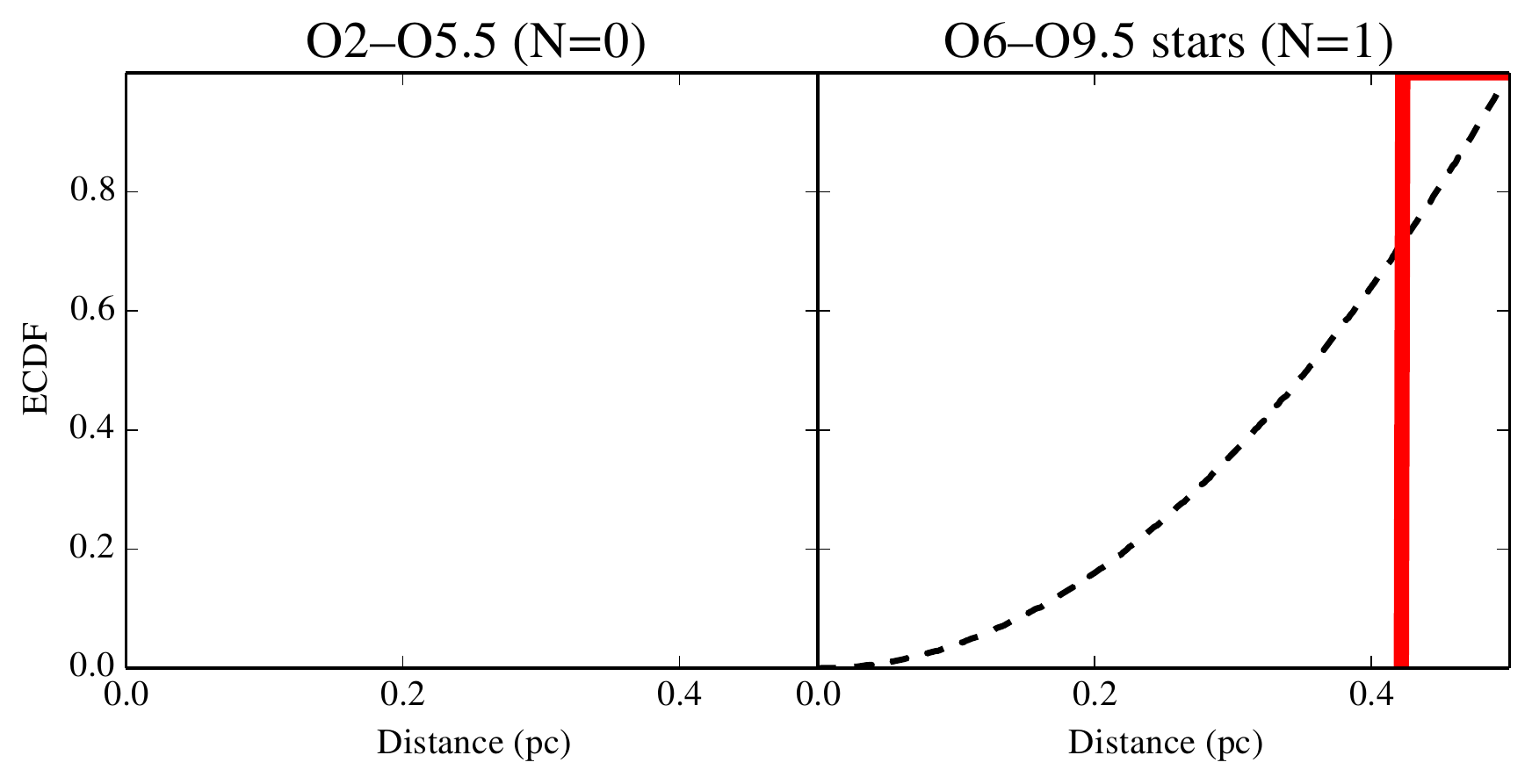}
\caption{NGC~6334. \ECDFcaption\ \ECDFcattwocaption}
\end{figure*}

\begin{figure*}[h]
\centering
\includegraphics[width=\ECDFwidth]{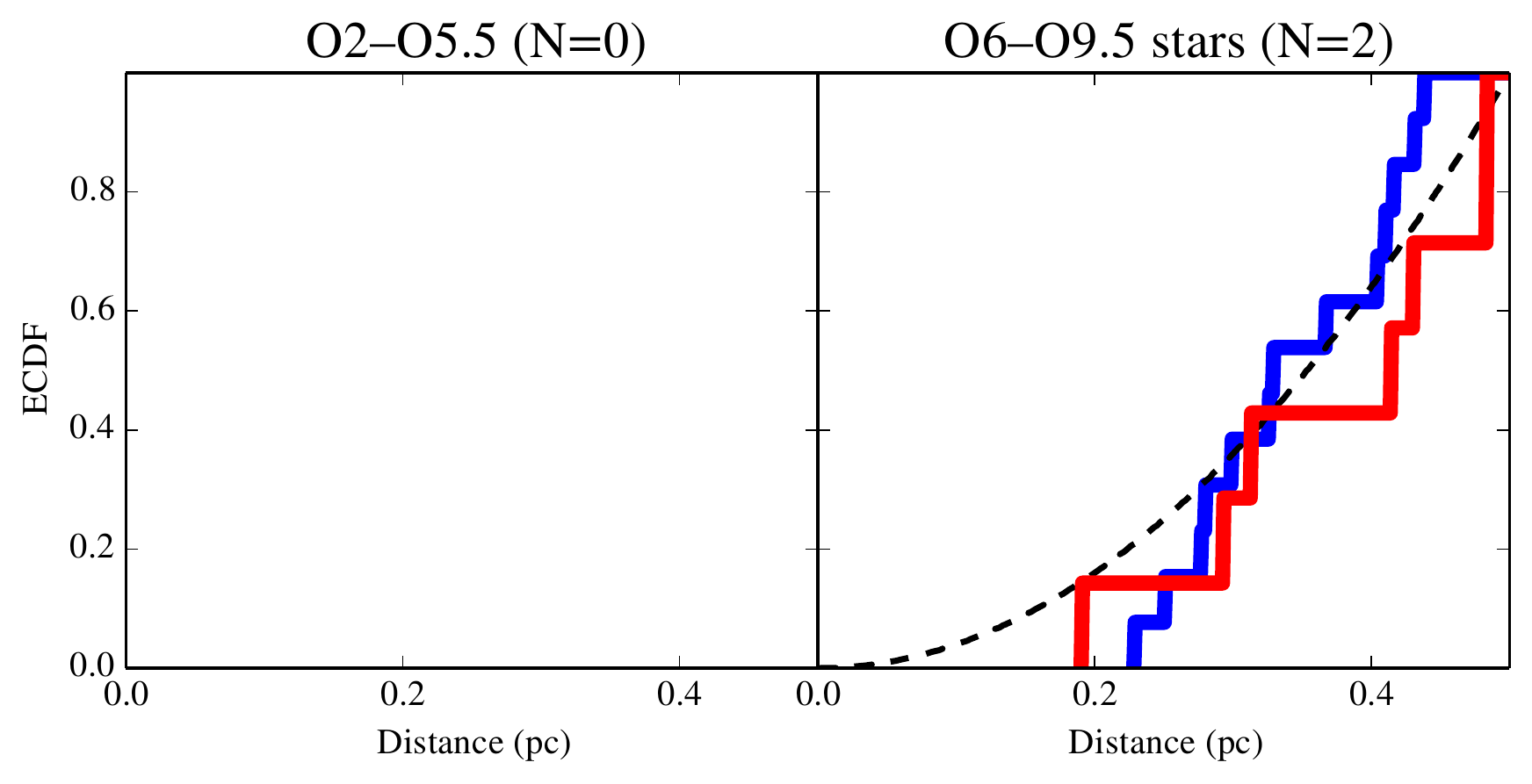}
\caption{W~3. \ECDFcaption\ \ECDFcattwocaption}
\end{figure*}

\begin{figure*}[h]
\centering
\includegraphics[width=\ECDFwidth]{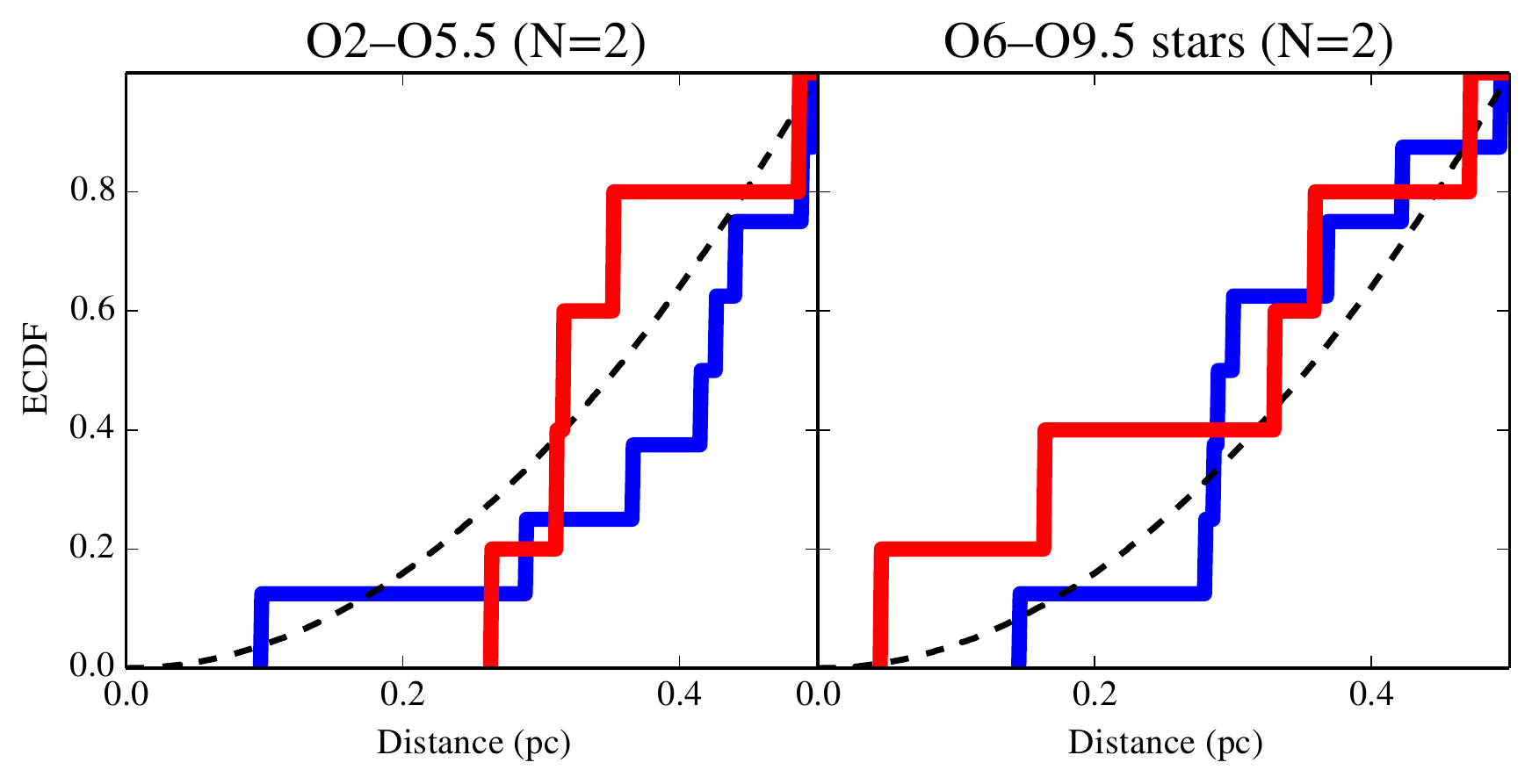}
\caption{W~4. \ECDFcaption\ \ECDFcattwocaption}
\label{fig:app2B}
\end{figure*}

\newpage

\input{bib.tex}
\end{document}

%% file: t1.tex
\begin{table}
\begin{center}
\caption{MYStIX region properties}
%\begin{tabular}{lccccc}
\label{table:regiondata}
\begin{tabular}{llrrrrr}
\hline
Region         &   \multicolumn{1}{c}{Dist}    &        \multicolumn{1}{c}{MLBF cut}       &      \multicolumn{1}{c}{$N_\mathrm{O}$}      & \multicolumn{1}{c}{{$N_\mathrm{D}$}}  & \multicolumn{1}{c}{{$N_\mathrm{ND}$}} & \multicolumn{1}{c}{{$\mathcal{M}_{50}$}} \\
               &    \multicolumn{1}{c}{pc}     &       \multicolumn{1}{c}{MJy~sr$^{-1}$}      &                          &                    &               &          {$M_\odot$}    \\
%               &    pc     &       MJy~sr$^{-1}$      &           \multicolumn{3}{c}{(after PAH region exclusion)}             \\
\hline
\multicolumn{7}{c}{\textit{No O stars (\S\ref{subsect:cat1})}}                                                                                 \\
DR 21                &   1.5     &       500                &            0             &        478         &           142      &  {0.3}  \\
Flame                &   0.414   &      5000                &            0             &        178         &            84      &  {0.8}  \\
NGC 2362             &   1.58    &        20                &            0             &         46         &           271      &  {0.3}  \\
RCW 36               &   0.7     &      3000                &            0             &        122         &            33      &  {0.2}  \\
RCW 38               &   1.7     &       700                &            0             &        109         &           134      &  {0.8}  \\
Trifid               &   2.7     &       225                &            0             &        163         &           140      &  {2.0}  \\
W 40                 &   0.5     &      2000                &            0             &        294         &            67      &  {0.1}  \\
\multicolumn{7}{c}{\textit{Low disk surface densities (\S\ref{subsect:cat2})}}                                                                 \\
Lagoon               &   1.3     &       425                &            2             &        444         &           634      &  {0.8}  \\
NGC 1893             &   3.6     &       200                &            5             &        287         &           205      &  {0.6}  \\
NGC 2264             &   0.914   &       400                &            1             &        533         &           401      &  {0.4} \\
NGC 3576             &   2.8     &       400                &            2             &        135         &           236      &  {2.0}  \\
NGC 6334             &   1.7     &      1250                &            1             &        396         &           317      &  {1.0}  \\
W 3                  &   2.04    &       300                &            2             &        241         &           283      &  {---}  \\
W 4                  &   2.04    &       100                &            4             &        153         &           186      &  {0.6}  \\
\multicolumn{7}{c}{\textit{High disk surface densities (\S\ref{subsect:cat3})}}                                                                \\
Carina         &   2.3     &       400                &           49             &        793         &          6007      &  {1.0} \\
Eagle          &   1.75    &       250                &            9             &        704         &           794      &  {1.7}  \\
M 17           &   2.0     &      3500                &           21             &        151         &           397      &  {1.0}  \\
NGC 6357       &   1.7     &      4000                &           10             &        518         &           675      &  {0.8}  \\
Orion          &   0.414   &     17500                &            1             &        626         &           853      &  {0.5}  \\
Rosette        &   1.33    &       200                &            6             &        599         &           749      &  {0.5}  \\
\hline                                                                                                                      
Total                &           &                          &          115             &       6970         &         12608      &             \\
\hline
\end{tabular}
\end{center}
\end{table}